\documentclass[12pt,nofootinbib,longbibliography]{revtex4-1}
\usepackage{amssymb}
\usepackage{amsmath}
\usepackage{graphicx}
\usepackage{color}
\usepackage[usenames,dvipsnames]{xcolor}
\newcommand{\scratch}[1]{}

\newcommand{\be}{\begin{equation}}
\newcommand{\ee}{\end{equation}}
\newcommand{\bea}{\begin{eqnarray}}
\newcommand{\eea}{\end{eqnarray}}
\newcommand{\half}{\f{1}{2}}
\newcommand{\f}[2]{\frac{#1}{#2}}
\newcommand{\ep}{\epsilon}
\newcommand{\pfrac}[2]{\left(\frac{#1}{#2}\right)}

\newcommand{\vT}{\vec T}

\newcommand{\varq}{q}
\newcommand{\ket}[1]{|#1\rangle}
\newcommand{\bra}[1]{\langle #1|}
\newcommand{\braket}[2]{\langle #1|#2\rangle}

\newcommand{\SLtC}{\rm{SL}(2,\mathbb{C})}

\newcommand{\Poincare}{Poincar\'e\ }
\renewcommand{\Im}{{\rm{\,Im\!}}}
\renewcommand{\Re}{{\rm{\,Re\!}}}

\begin{document}

\title{On the Theory of Continuous-Spin Particles: \\ Wavefunctions and Soft-Factor Scattering Amplitudes}

\author{Philip Schuster}
\email{pschuster@perimeterinstitute.ca}
\affiliation{Perimeter Institute for Theoretical Physics,
Ontario, Canada, N2L 2Y5 }

\author{Natalia Toro}
\email{ntoro@perimeterinstitute.ca}
\affiliation{Perimeter Institute for Theoretical Physics,
Ontario, Canada, N2L 2Y5 }
\date{\today}

\begin{abstract}
The most general massless particles allowed by Poincar\'e-invariance are ``continuous-spin'' particles (CSPs) characterized by a scale $\rho$, which at $\rho=0$ reduce to familiar helicity particles.  Though known long-range forces are adequately modeled using helicity particles, it is not known whether CSPs can also mediate long-range forces or what consequences such forces might have.  
We present sharp evidence for consistent interactions of CSPs with matter: new CSP equations of motion, wavefunctions, and covariant radiation amplitudes.
In companion papers, we use these results to resolve old puzzles concerning CSP thermodynamics and exhibit a striking correspondence limit where CSP amplitudes approach helicity-0, 1 or 2 amplitudes.
\end{abstract}

\maketitle

\newpage
\tableofcontents
\newpage

%%%%%%%%%%%%
\section{Introduction}
\begin{figure}[!htbp]
\includegraphics[width=\columnwidth]{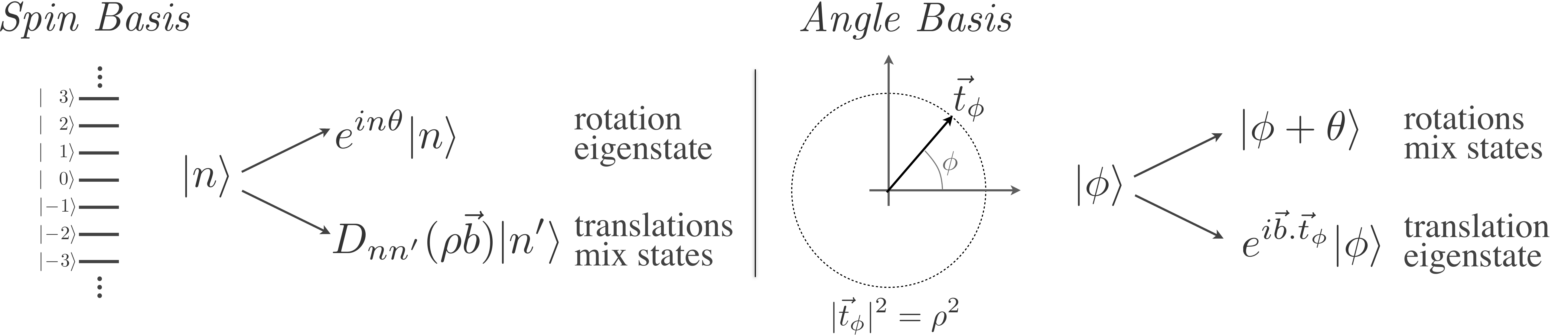}
\caption{The figure summarizes the Little Group (LG) transformation of massless particle states (see \S\ref{sec:CSR_reps}).  
Particle types are characterized by a scale $\rho$. Basis states may be labeled by a tower of integer or half-integer spins, 
or equivalently by angles on a circle.  The two bases are related by Fourier transform. 
The LG has the structure of the isometries of the Euclidean 2-plane, or $ISO(2)$. 
The spin basis diagonalizes LG rotations, while the angle basis diagonalizes LG translations.
Lorentz boosts induce LG translations (and rotations), which mix states in the spin basis. 
The scale $\rho$ controls the amount of mixing under boosts, much like the combination 
$m\times S$ for a spin-$S$ massive particle.
When $\rho=0$, spin labels become Lorentz-invariant helicities. \label{fig:spinAngle}}
\end{figure}

Massless particles are typically classified by their helicity, the eigenvalue of the operator $\hat{h}\equiv\vec{\bf k}.\vec{\bf J}/|{\bf k}|$, 
where $\vec{\bf k}$ is the particle's three-momentum and $\vec{\bf J}$ the angular momentum operator.  
This operator is plainly not boost-invariant, so why do we speak of boost-invariant helicities at all? 
Wigner's 1939 classification of particles consistent with Poincar\'e symmetry showed that indeed, Lorentz-invariance of helicity is not generic \cite{Wigner:1939cj}.  
Wigner found a new Lorentz covariant massless particle type , the ``continuous-spin'' particle (CSP), labeled in 3+1 dimensions by a spin-scale $\rho$ with units of mass.
Its single-particle states can still be labeled by integer $\hat h$ eigenvalues, \footnote{A second continuous-spin representation in 3+1 dimensions, which we do not consider here, takes on half-integer $\hat h$ eigenvalues. Supersymmetric and higher-dimensional continuous-spin representations also exist.} 
but the states of different spins mix under Lorentz boosts, much like massive spin states.  The degree of mixing is controlled by $\rho$, and in the $\rho=0$ limit the CSP factorizes into a tower of $\hat h$ eigenstates that do not mix under Lorentz transformations -- only then can we speak of Lorentz-invariant helicities. 
States can equivalently be labeled (via Fourier transform) by an angle in [0,$2\pi$), giving rise to the name \emph{continuous-spin} (see Figure \ref{fig:spinAngle}).

Very little is known about CSP dynamics, and in particular whether or how CSPs can interact with matter --- a rather striking omission 
in our understanding of massless particles and long-range physics. 
This is the first of a series of papers in which we present a self-contained introduction to CSPs,
evidence that they can interact consistently with matter, evidence for a correspondence of CSP interactions
with those of familiar helicity particles \cite{Schuster:2013vpr}, certain physical consequences of this correspondence \cite{SchusterToro:ph}, 
and the formulation of a new gauge field-theory description of CSPs coupled to a background current \cite{Schuster:2013pta}.
%that relates at $\rho=0$ to conventional descriptions of arbitrary-helicity massless particles \cite{Schuster:2013pta}. 
Our findings expose key structures that any consistent CSP theory must reproduce and suggest new lines of attack towards determining  
 whether CSPs can couple to gravity and to each other and understanding the locality properties of CSP interactions with matter. 

Here we review the kinematics of CSPs and report two classes of discoveries. 
We present new wave equations and covariant wavefunctions for CSPs, which differ qualitatively from the previous state of the art based on Wigner's wave equations \cite{Wigner:1939cj,Wigner:1947,Bargmann:1948ck}.
A new family of ``smooth'' wavefunctions is particularly useful for constructing scattering amplitudes. 
Unlike Wigner's equations, our new equations of motion make direct contact at $\rho=0$ with familiar formalisms for describing massless particles.
Building on these findings, we obtain ``soft factors'' useful for constructing CSP scattering amplitudes. 
Any theory of interacting CSPs --- if one exists --- must recover amplitudes of this form in the soft limit.  
These soft factors are analytic functions of momentum, but with an isolated essential singularity at soft and/or collinear configurations. 
The singularity is much better behaved than any finite-order pole, and does not appear to correspond to new degrees of freedom.
These soft factors can be sewn together to build ansatz amplitudes that are appropriately bounded, unitary, analytic functions of momentum \cite{Schuster:2013vpr}.  

The existence of covariant soft emission amplitudes for CSPs is striking and nontrivial evidence that they can consistently interact.
Indeed, no covariant soft factors exist for high-helicity particles \cite{Weinberg:1964ev,Weinberg:1964ew}.
Moreover, the structure of CSP soft factors suggests that interacting CSPs would mediate long-range forces!
This possibility is particularly exciting in light of evidence for a \emph{helicity correspondence}: 
CSPs with energy large compared to $\rho$ (or $\rho v$ for a non-relativistic emitter) 
behave to a good approximation like definite-helicity particles accompanied by a tower of very weakly interacting states \cite{Schuster:2013vpr}.  
The conjecture that general CSP amplitudes exhibit helicity correspondence raises new theoretical and phenomenological possibilities, developed in 
more detail in \cite{Schuster:2013vpr, SchusterToro:ph}.

It is striking that continuous-spin particles, which fit so poorly into our present theoretical understanding, may be among the few excitations capable of mediating long-range forces.  Whether or not CSPs are realized in Nature, perhaps this signals a conceptual gap in our understanding of infrared physics.  Though history took a different course, the physics of flat-space helicity-2 excitations would have sufficed to develop General Relativity \cite{Feynman:1996kb,Weinberg:1965rz}.  Likewise, the flat-space physics of CSPs may point towards a more powerful formalism that describes CSPs, gauge theories, and General Relativity on an even footing.   

We briefly comment on two classes of objections to CSPs raised in the literature.
The most physical concern, due to Wigner \cite{Wigner:1963}, is that theories with CSPs have infinitely many degrees of 
freedom and hence infinite heat capacity per unit volume.
Indeed, if all CSP states thermalized democratically and rapidly enough, it would lead to rapid supercooling of all thermal systems.  
However, Lorentz invariance requires a hierarchical coupling structure in soft factors (see \S\ref{sec:WFandSoftFactor}), which is expected to persist in general amplitudes with helicity correspondence \cite{Schuster:2013vpr}.  This characteristic structure \emph{guarantees} that in realistic approximately thermal systems, the ``CSP bath'' does not thermalize --- even for microscopic $\rho^{-1}$ \cite{SchusterToro:ph}!
A second concern dates from the 1970s, when 
several groups built covariant CSP fields and found obstructions to either canonically quantizing these fields or building a local Hamiltonian \cite{Iverson:1971hq,Abbott:1976bb,Hirata:1977ss}. All but one of these authors used Wigner's singular wavefunctions, which do not satisfy familiar wave equations  when $\rho=0$.  Moreover, all three assumed both Lorentz-covariance of the fields and a one-to-one correspondence with single-particle states --- a strategy that would have failed to construct quantum electrodynamics, or any other gauge theory!
In \cite{Schuster:2013pta}, we present a gauge field theory for CSPs with local field operators that satisfy 
familiar quantization conditions. This theory is equivalent to known gauge theories when $\rho=0$,
though it is unclear whether interactions with matter can be manifestly local. 
The earlier difficulties are directly related to overly restrictive assumptions about how to implement Lorentz covariance and the use of Wigner's singular wavefunctions.

In Section \ref{sec:CSR_reps}, we review Wigner's Little Group classification of one-particle states in Poincar\'e-invariant theories \cite{Wigner:1939cj}, emphasizing CSPs.
In Section \ref{sec:wavefunctions}, we derive the most general Little Group and Lorentz covariant wavefunctions that describe
CSPs, many of which are new. We then show how a subset of these wavefunctions
solve the Wigner equations, while the broader class form bases of solutions to entirely new wave equations. 
In Section \ref{sec:WFandSoftFactor}, we introduce CSP soft factors, explain how they can be used to build a limited class
of candidate CSP scattering amplitudes, and discuss basic analytic properties of these amplitudes. 
In Section \ref{sec:conclusion}, we summarize our findings and describe open problems regarding CSP interactions that should be resolved.

%Citations:
%\cite{Schuster:2013vpr,Schuster:2013pta}
%\cite{Wigner:1939cj,Bargmann:1946me} 
%\cite{Weinberg:1964ev,Weinberg:1964ew,Weinberg:1965rz}
%\cite{Benincasa:2007xk,Porrati:2012rd}
%\cite{Yngvason:1970fy,Iverson:1971hq,Chakrabarti:1971rz,Abbott:1976bb,Hirata:1977ss}

%%%%%%%%%%%%%%%%%%%%%%%%%%%%%%%%
\section{Poincar\'e-Covariance and Continuous-Spin Particles in 3+1 Dimensions}\label{sec:CSR_reps}
Wigner has classified all possible one-particle states in Lorentz-invariant theories by requiring that they transform as unitary irreducible representations (irreps) of 
the \Poincare group~\cite{Wigner:1939cj}.  The Poincar\'e-transformation of single-particle states dictates the Lorentz transformation properties of scattering amplitudes. 
Weinberg used this connection to derive constraints on high-helicity particles' interactions \cite{Weinberg:1964ev,Weinberg:1964ew,Weinberg:1965rz} independent of any specific field theory; we apply the same approach to continuous-spin particles in this paper.  

Though most readers are familiar with Wigner's ``Little Group'' construction in 3+1 dimensions (see e.g. \cite{WeinbergQFT}), we review it here to highlight aspects that are not emphasized in textbook treatments, but will prove useful in treating continuous-spin particles.  
In particular we emphasize a covariant formulation, where the physical interpretation of the ``standard Lorentz transformation'' 
appearing in the Little Group construction is that it defines a choice of coordinate system for the Little Group associated with each four-momentum.
Much of this material appears, with different emphasis, in \cite{Iverson:1971hq,Boels:2009bv}.

To prime this discussion, we highlight puzzles in the usual definitions of spin (helicity) for massive (massless) particles.  
Though spin $S$ and helicity $h$ are Lorentz-invariant, they are usually defined as eigenvalues of operators that are not manifestly Lorentz-invariant:
\bea
{\vec {\bf J}}^2 \ket{\psi_S} &=& S (S+1) \ket{\psi_S}, \label{spinRest}\\
\vec {\bf k}.\vec{\bf J}/|{\bf k}| \ket{\psi_h}& = & \hat h \ket{\psi_h},\label{helicity3V}
\eea
where $\vec{\bf J}$ is the 3-vector of rotation generators and $\vec{\bf k}$ the massless particle's 3-momentum.  Can these formulas be made Lorentz-invariant? 

These two apparently similar puzzles have very different resolutions.  We will derive an improved, manifestly covariant form of \eqref{spinRest} from the covariant 
formulation of the Little Group in \S\ref{sec:LG}. For any four-momentum $k^\mu$, the Little Group is generated by the three independent components of
\be
w^\mu \equiv  \half \epsilon^{\mu\nu\rho\sigma} k_{\nu} J_{\rho\sigma} \label{wkdefINTRO}
\ee
which clearly reduces to $m(0,\vec{\bf J})$ for a particle at rest.  Spin can be more covariantly defined by 
\be
w^2 \ket{\psi_S} = - m^2 S(S+1) \ket{\psi_S}\label{massivew2}
\ee
where $m^2=k^2$ (using the mostly-negative metric).
Physically, $\sqrt{-w^2}/E$ characterizes the decoupling of adjacent spin states at energy $E$ --- hence the well-known helicity approximation for high-energy scattering of spin-$1/2$ or spin-1 particles.  The $S$-scaling is less familiar, but intuitively if all states of a spin-$S$ particle are easily mixed at energy $E\sim m$, nearest-neighbor states must begin to mix significantly at $E\sim mS$.

\begin{figure}[!htbp]
\includegraphics[width=0.55\columnwidth]{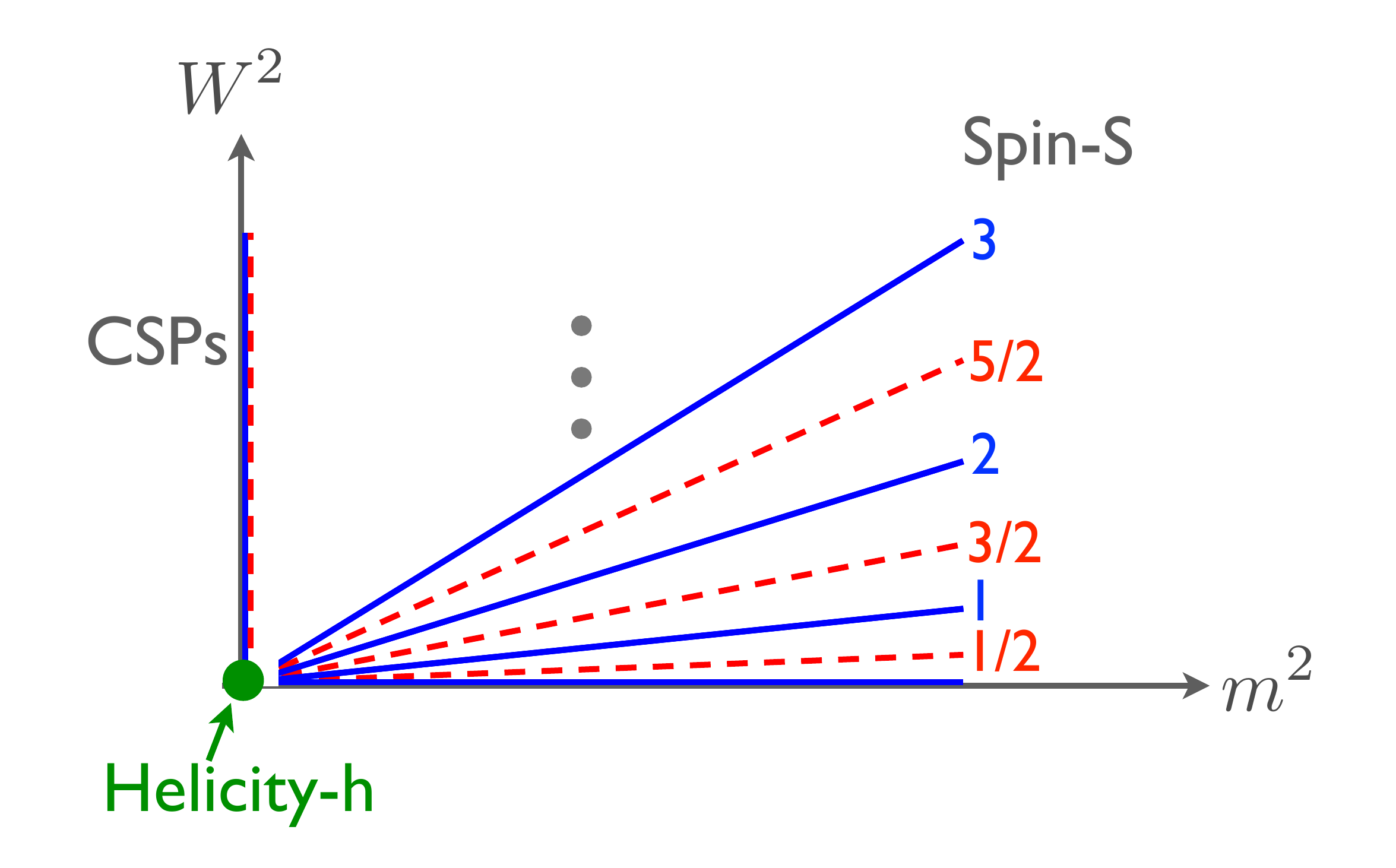}
\caption{Irreducible representations of the \Poincare group are labelled by the square of the Pauli-Lubanski vector operator $W^2$, 
and the square of the momentum operator $P^2=m^2$. The plot illustrates the relation between $W^2$ (y-axis) and $P^2$ (x-axis)
imposed by the structure of the Little Group. As $P^2\rightarrow 0$, the massive branches approach the 
continuous-spin branch as $S(S+1)\rightarrow \rho^2/P^2$.  Continuous-spin representations decompose into helicity representations for $\rho \rightarrow 0$ along the $m^2=0$ slice.  The helicity-$h$ states $(h\geq 1)$ are not continuously related to the massive branches. 
\label{fig:reps}}
\end{figure}

In contrast, the helicity operator in \eqref{helicity3V} cannot be made manifestly Lorentz-invariant at all --- a first hint that Lorentz invariance alone does 
not guarantee the boost-invariance of massless particles' helicities.  
Indeed, \eqref{massivew2} has a massless generalization 
\be
w^2 \ket{\psi_\rho} = - \rho^2 \ket{\psi_\rho},
\ee
where $\rho$ is an unconstrained dimensionful ``spin scale''.  Only when $w^2=0$ is helicity (as defined by \eqref{helicity3V})
a boost-invariant quantum number \footnote{In this case, a given helicity $h$ is singled out by the requirement $(w^{\mu}-hP^{\mu})\ket{\psi_h}=0$.}.  The ``continuous-spin'' representations with $w^2\neq 0$ contain an infinite tower of $\hat h$-eigenstates with integer-spaced eigenvalues.  The intuition mentioned above for massive particles, that $\rho/E$ controls the mixing of adjacent spin states, continues to apply.  
The relationship of $w^2$ and $p^2$ for different representations of the \Poincare group 
is summarized in Figure \ref{fig:reps}.

\subsection{\Poincare Action on One-Particle States}
To begin, we consider the Lorentz transformation of single-particle states in irreps of the \Poincare group.  Because the translations 
$P^\mu$ mutually commute, states can be labeled by a $c$-number momentum eigenvalue $k^\mu$ and by internal labels  
$a$ whose detailed form we will constrain later.  We may write the action of a Lorentz transformation $\Lambda$ on each state as 
\be
U(\Lambda) \ket{k,a} =\sum_{a'} D(\Lambda,k)_{aa'} \ket{\Lambda k, a'},\label{LorentzTransf}
\ee
where the transformation matrix $D$ must be unitary with respect to the norm
\be
\braket{k,a }{k', a' } = 2 k^0 \delta^{(3)}(k-k') \delta_{aa'} \label{norm}. 
\ee
These formulas apply to discrete labels $a$; if $a$ is continuous, then $D$ is a transformation \emph{function},  
the sum over $a'$ in \eqref{LorentzTransf} becomes an integral, and $\delta_{aa'}$ in \eqref{norm} goes to $\delta(a-a')$.

We will focus first on two special types of Lorentz transformations --- those that leave the momentum $k$ invariant but change $a$, and 
those that change $k$ but leave $a$ unchanged (the former are fixed by group theory, while the latter must be chosen by convention). Any Lorentz transformation can be decomposed as a product of transformations of these two types.  
In the next section, we will solve for the subgroup of Lorentz transformations $W$ that leave $k^\mu$ invariant 
(the \emph{Little Group} of $k$, $LG_k$).  By \eqref{LorentzTransf}, these act on states as
\be
U(W)  \ket{k,a} =\sum_{a'} D(W,k)_{aa'}  \ket{k, a'} \qquad \mbox{(for $W\in LG_{k}$)}, \label{LGAction}
\ee
which means that the $D$'s must furnish a unitary  representation of $LG_k$ on the labels $a$.  These representations will be classified in the next two sections.

Whereas the Little Group is fully determined by $k$, the choice of Lorentz transformations that take $\ket{k,a}$ to $\ket{k',a}$ (for $k^2=k'^2$) 
is dictated purely by convention.  It is standard to construct all states in a given \Poincare irrep from the states at \emph{fixed} reference 
momentum $\bar k^\mu$, as follows.  For each $k$ we choose a ``standard Lorentz transformation'' $B_k$ such that
$(B_k)^\mu_\nu \bar k^\nu =k^{\mu}$, for which we \emph{define} 
\be
D(B_k,k)_{a a'} \equiv \delta_{a a'}. \label{stdAction}
\ee
The action of any Lorentz transformation $\Lambda$ on one-particle states is then  determined by group composition of   \eqref{LGAction} and \eqref{stdAction} to be 
\be
U(\Lambda)\ket{k,a} = D(W_{\Lambda,k},k)_{a a'} \ket{\Lambda k, a'} 
\quad \mbox{with} \quad W_{\Lambda,k} \equiv  B_{\Lambda k}^{-1} \Lambda B_k \in LG_{\bar k},
\label{particleDef}.  
\ee
So far, this construction is entirely conventional, but it somewhat obscures the
physical significance of the choice of $B_k$. We will return to this point after identifying the Little Group for a general $k^\mu$ in the next section.

\subsection{The Little Groups for Massive and Massless Particles}\label{sec:LG}
To classify Lorentz transformations that leave a momentum $k_\mu$ invariant (the Little Group $LG_k$), we first 
parametrize the Lorentz generators as $G_Y = \epsilon^{\mu\nu\rho\sigma} Y_{\mu\nu} J_{\rho\sigma}$ with $Y_{\mu\nu}$ 
an arbitrary antisymmetric tensor.  The action of $G$ on $k^\alpha$ is given by
\be
(G_Y)^\beta_\alpha k^\alpha =  \epsilon^{\mu\nu\rho\beta} Y_{\mu\nu} k_\rho,
\ee
which vanishes if and only if $Y_{\mu\nu} = y_{[\mu} k_{\nu]}$ for some $y_{\mu}$.  
Therefore $LG_k$ is generated by the independent components of  
\be
w^\mu \equiv  \half \epsilon^{\mu\nu\rho\sigma} k_{\nu} J_{\rho\sigma}. \label{wkdef}
\ee
There are only three independent components of $w^\mu$ because $w.k=0$ --- these are the three little-group generators 
(in $(1,D-1)$ dimensions $w^\mu$ generalizes to an orthogonal, antisymmetric $D-3$-tensor worth of generators). 
We note the close relationship of $w^\mu$ to the Pauli-Lubanski pseudo-vector operator 
\be
W^\mu \equiv \half \epsilon^{\mu\nu\rho\sigma} P_{\nu} J_{\rho\sigma},\label{Wdef},
\ee
with $P_{\nu}$ replaced by its $c$-number eigenvalue $k^\mu$.   Indeed, $W^2$ can be shown to commute 
with all \Poincare generators, and therefore its action on any irreducible representation of the \Poincare group is proportional to the 
identity.  We can therefore classify representations by the two invariants
\be
P^2 = m^2 \qquad W^2 = -\rho^2,
\ee
where $\rho$ has units of linear momentum.   The group structure of the Little Group can be 
inferred from the Pauli-Lubanski pseudo-vector's commutation relations 
\bea
\left[W^\mu,W^\nu\right] &=& - i \epsilon^{\mu\nu\rho\sigma} W_\rho P_\sigma. \label{WWcommutator}
\eea

\textbf{Massive Particles:} In a mass-$m$ particle's rest frame, $w^\mu$ reduces to $w^0 = 0$, $w^i = m J^i$, implying an $SO(3)$ Little Group structure and the resulting quantization condition 
\be
\f{1}{m^2} w^2 \ket{\psi_j} = - S (S+1) \ket{\psi_j} \label{spinCovariant}
\ee
where $w^2=w^{\mu}w_{\mu}$.
This relation provides a manifestly Poincar\'e-invariant definition of a massive particle's spin, generalizing \eqref{spinRest},
and forces $W^2$  for \emph{massive} particles to take on discrete values: 
 \bea W^2 &\equiv& -\rho^2=-m^2c^2S(S+1), \nonumber \\
&=& - \left( \frac{mc}{\hbar} \right)^2 \hbar^2S(S+1).
\eea
We have re-instated units of $\hbar$ and $c$ to emphasize that although $S$ is quantum-mechanical, 
$W^2$ (and $\rho$) is also entirely well-defined in a classical $\hbar\rightarrow 0$ limit. 
In Appendix \ref{App:LG}, we explicitly construct the Little Group and derive commutation relations for a general massive 
(timelike) momentum $k^\mu$, in a way that connects smoothly to the massless case.  
 
\begin{figure}[!htbp]
\includegraphics[width=0.65\columnwidth]{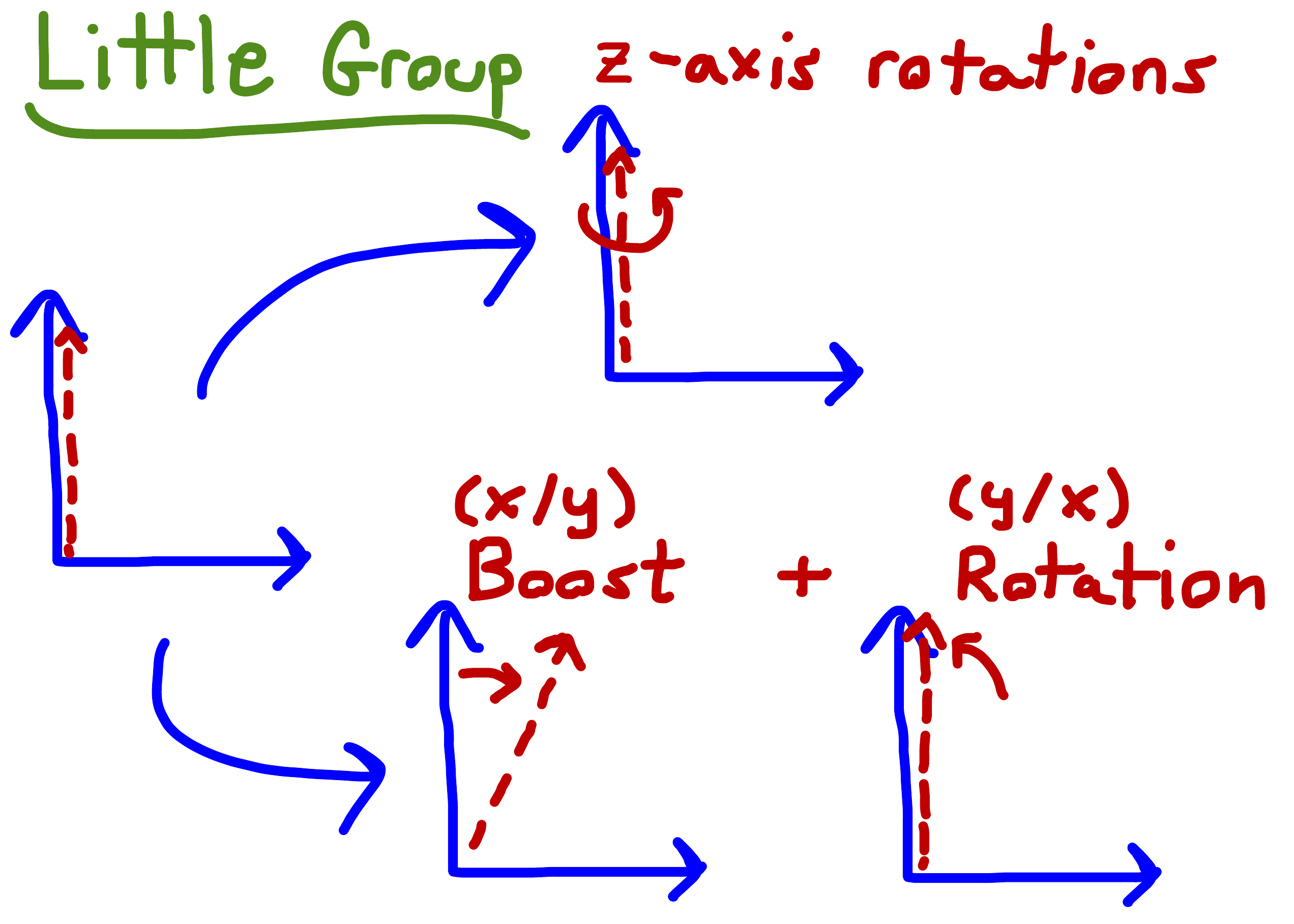}
\caption{Three Lorentz transformations that leave a null vector $k^\mu$ invariant are illustrated above. These form a basis for the Little Group of $k^\mu$.  
For the choice $\bar k^\mu=(\omega,0,0,\omega)$, these are $J_{12}$, $(J_{32}+J_{02})$, and $(J_{31}+J_{01})$.
In addition to the obvious rotation about the 3-momentum axis (top), two combinations of rotations and transverse boosts (bottom)
also leave $\bar k$ invaraint. These three transformations form the group of isometries of the Euclidean 2-plane.\label{fig:m0LGgens}}
\end{figure}

\textbf{Massless Particles:} In the case of null momentum ($k^2=0$), the constraint $w.k=0$ motivates a decomposition of $w^\mu$ into a ``rotation'' component proportional to $k^\mu$ and two ``translation'' components along polarization directions $\epsilon_{1,2}^\mu$ with $\epsilon_{1,2}.k=0$ (see \cite{Heinonen:2012km} for a similar treatment): 
\be
w^\mu = - k^\mu R + \epsilon_1^\mu T_1 + \epsilon_2^\mu T_2. \label{wdecomp}
\ee
For example, with $\bar k^\mu=(\omega,0,0,\omega)$, we can expand \eqref{wkdef} in components as
\be
w^\mu = - \bar k^\mu J_{12} + \hat e_x^\mu \left( \omega (J_{32}+J_{02}) \right) + \hat e_y^\mu \left( -  \omega (J_{31}+J_{01}) \right). \label{wframe}
\ee
Note that the components $J_{12}$, $(J_{32}+J_{02})$, and $(J_{31}+J_{01})$ are the generators for Lorentz transformations
that leave $\bar k$ invariant -- this is illustrated in Figure \ref{fig:m0LGgens}. 
The components $R$ and $T_{1,2}$ can be extracted as
\be
R=q.w \qquad T_{1,2} = \epsilon_{1,2}.w, \label{Wm0components}
\ee
where $q$ is the unique vector satisfying
\be
q^2=0, \quad p.q=1, \quad q.\epsilon_{1,2}=0. \label{eq:qdef}
\ee

From \eqref{WWcommutator}, we find commutators 
\be
[R,T_{1,2}] = \pm i  T_{2,1}, \qquad [T_1,T_2] = 0,
\ee
so $LG_k$ has the structure of $ISO(2)$, the isometry group of a Euclidean plane (hence the names assigned to the
generators).  
It is useful to define a two-vector of ``little-group translations'' $\vec T=(T_1,T_2)$, and also to group these into conjugate 
raising and lowering generators $T_\pm = T_1 \pm i T_2$ with commutators $[R,T_\pm] = \pm T_\pm$.  These act on vectors as 
\be
{R^\mu}_{\nu} = \epsilon_-^\mu {\epsilon_+}_\nu - \epsilon_+^\mu {\epsilon_-}_\nu,
\quad 
{(T_\pm)^\mu}_\nu = \pm \sqrt{2} \left( \epsilon_\pm^\mu k_\nu -k^\mu {\epsilon_\pm}_\nu \right), \label{LGgenAction}
\ee
where $\epsilon_\pm = (\epsilon_1 \pm i \epsilon_2)/\sqrt{2}$, which follow from \eqref{wkdef} and \eqref{Wm0components} (see also Appendix \ref{App:LG}).  
As in the massive case, the invariant 
\be
W^2 = w^2 = - \vec{T}^2 =  - T_+ T_- \label{w2massless}
\ee
can be used to classify representations, though here (unlike the massive case) the group structure does not imply quantization of $w^2$.
We stress that, because $k$ is null, the rotation generator $R$ drops out of the expression for $W^2$!

For future use in constructing covariant wavefunctions, we introduce a canonical decomposition of any LG element as 
\be
W(\theta,\beta)\equiv e^{\frac{i}{\sqrt{2}}(\beta T_{-} + \beta^* T_+)} e^{-i\theta R},\label{LGelem}
\ee
where $\theta \in [0,2\pi)$ and $\beta$ is complex and has dimensions of length, as the translation generators have units of mass. 
This little group element transforms the reference frame vectors $\epsilon_\pm(k)$ and $q(k)$ defined by \eqref{eq:qdef} as 
\bea
%W(\theta,\beta)^{\mu}_{\nu}\eta^{\nu}&\equiv& (e^{\frac{i}{\sqrt{2}}\beta w_{-} + \frac{i}{\sqrt{2}}\beta^* w_{+}} e^{-i\theta w_{r}} )^{\mu}_{\nu}\eta^{\nu} ,\\
W(\theta,\beta)^{\mu}_{\nu}\epsilon_{+}^{\nu}(k)&=& e^{-i\theta}\left( \epsilon^{\mu}_{+}(k)-i\beta k^{\mu} \right) \label{eq:ALGonEpsilon}\\
W(\theta,\beta)^{\mu}_{\nu}\epsilon_{-}^{\nu}(k)&=& e^{i\theta}\left( \epsilon^{\mu}_{-}(k) +i\beta^* k^{\mu} \right) \\
W(\theta,\beta)^{\mu}_{\nu}q^{\nu}(k) &=& q^{\mu}+k.q (i\beta^*\epsilon^{\mu}_{+} -i\beta\epsilon^{\mu}_{-}+|\beta|^2k^{\mu}) \label{eq:BLGonEpsilon}.
\eea
We should stress that these are Little-Group actions on $\epsilon_\pm^\mu(k)$, \emph{not} Lorentz transformations!  We have defined $\epsilon$ to be a function of $k$, so that Lorentz-transformations take $\epsilon_\pm^\mu(k)$ to $\epsilon_\pm^\mu(\Lambda k)$ --- a non-tensorial transformation, except for the special cases $\Lambda = B_{\Lambda k} B_k^{-1}$.
It is clear that this whole construction could be repeated with a different choice of $\epsilon_{1,2}$ for the same momentum $k^\mu$.  
This procedure would identify different generators $T'_{\pm}$ and $R'$ for the same Little Group.  This ambiguity is fixed by choosing $\epsilon_\pm$ at a standard momentum $\bar k$, then choosing for each $k$ a standard Lorentz transformation $B_k$ that maps $\bar k$ to $k$.  Choosing for example 
\be
\bar k = (\omega,0,0,\omega), \quad \epsilon_\pm^\mu(\bar k)  \equiv (0,1,\pm i, 0)/\sqrt{2},
\ee
the boost $B_k$ will take $\bar k^\mu$ to $k^\mu$ and $\epsilon_\pm(\bar k)$ to new complex null $\epsilon_\pm^\mu(k)$, which will always satisfy 
\be
(\epsilon_+(k))^* = \epsilon_-(k), \quad  \epsilon_+(k).\epsilon_-(k) = -1 , \quad \epsilon_\pm(k).k= 0 \label{simpleFrame}
\ee
and therefore are a consistent choice of frame at $k$.  Having chosen $\epsilon_\pm$ for each $k$, the relation \eqref{Wm0components} specifies the $T_\pm$ and $R$ generators. 

This connection clarifies the physical significance of a choice of standard boost $B_k$. 
In \eqref{particleDef}, it is natural to think of the $a$'s in $\ket{k,a}$ as labelling states in an $LG_k$ representation, while the $a$'s in $\ket{k',a}$ label states 
in an $LG_{k'}$ representation.  These two Little Groups, though similar, are distinctly embedded in the Lorentz Group so there is no 
Lorentz-invariant notion of the ``same'' LG state at different momenta.  To unambiguously relate Little Group \emph{states} with momentum $k$ to 
those at $k'$, we must relate our coordinate systems (frames) for the \emph{groups} $LG_k$ and $LG_{k'}$.  This is precisely what a choice of $B_k$ does ---
the choice of frame $\epsilon_{\pm}(\bar k)$ for $LG_{\bar k}$ and of $B_k$ selects a frame $\epsilon_{\pm}(k)\equiv B_k\epsilon_\pm(\bar k)$ for $LG_{k}$.  A different ``standard transformation'' $B'_k = W B_k$ with $W \in LG_k$ would induce a different coordinate system for $LG_{k}$, and hence a different labelling of states.
Alternately, because a Lorentz transformation is fully specified by its action on $\bar k^\mu$ and $\epsilon_\pm^\mu(\bar k)$, specifying 
$\epsilon_\pm^\mu(k)$ implicitly defines $B_k$.  Throughout the remainder of this paper, it will be convenient to leave the choice of $B_k$'s implicit and instead specify frame vectors for each $k$.

In particular physical regimes, different choices of $B_k$ are useful.  For example, it is possible to choose $B_k$'s for a massive particle such 
that for any $k$ and any pure rotation $R$, the induced Little Group transformation $W_{R,k}$ as defined in \eqref{particleDef} is $R$ itself.  
A particularly useful $B_k$ for working with continuous-spin particles is 
\be
B_k = e^{i \phi J_z}   e^{i \theta J_y}  e^{i \log(|{\bf k}|/\omega) K_z},
\ee
which maps $\bar k$ to $k = |{\bf k}| (1, \sin\theta\cos\phi, \sin\theta \sin\phi, \cos\theta)$ and $\epsilon_\pm(\bar k)$ to $\epsilon_\pm(k)$ that satisfy 
\be
\epsilon^0_\pm(k)= 0 \quad (\vec \epsilon_+ + \vec \epsilon_- ).\vec {\bar k} = 0
\ee
in addition to \eqref{simpleFrame}.  The analogue of this construction for the massive LG is given in Appendix \ref{App:LG}.  

\subsection{ Helicity and Continuous-Spin Representations of $ISO(2)$}
There are two very different classes of unitary irreps of the LG.
One-dimensional \textbf{helicity representations}, labeled by
a helicity $h$, consist of a single state on which the non-compact
translation generators $\vT$ act trivially and the rotation $R$ acts as a phase:
\be
W[\theta,\beta] \ket{k,h} = e^{i h \theta} \ket{k, h}.
\ee
(Here and in the following, we will often write $W|k,\phi \rangle$ and $\langle k,\phi| W^\dagger$ as shorthand in place of $U(W)|k,\phi\rangle$ and  $\langle k,\phi| U(W)^\dagger$ for the unitary action of $W$ on in- and out- states.)
The requirement of periodicity under rotations by $4\pi$ restricts 
$h$ to  be integer or half-integer.  By \eqref{w2massless}, all helicity representations have  $\rho^2 =
-W^2 = 0$.  These are the only \emph{finite-dimensional} representations of the massless Little Group\footnote{Similarly, in higher dimensions the only finite-dimensional representations are those on which the non-compact generators of $ISO(D-2)$ act trivially.  In 2+1 dimensions, the situation is quite different: the massless little group $ISO(1)$ has no compact generators, so there is no analogue to the helicity representations, while the ``continuous-spin-like'' representations with $\rho\neq 0$ are one-dimensional}.

\textbf{Continuous-spin representations}, labeled by an arbitrary
positive $\rho^2$ of mass-dimension 2, have a countable tower of states on which all Little Group generators act non-trivially.  
These can be described in two alternate bases: $R$-eigenstates (the ``spin basis'' labeled by an arbitrary integer or half-integer), 
or simultaneous eigenstates of $T_{1,2}$ (the ``angle basis'' labeled by an angle in $[0,2\pi)$).  
The labelling of states and Little Group actions in each basis are summarized in Figure \ref{fig:spinAngle}. 
The spin basis naturally makes contact with helicity and high-spin massive 
particles\footnote{At the level of group theory, the continuous-spin representations are just the $m\rightarrow 0$, $j\rightarrow \infty$,
$m\times j \rightarrow \rho$ limit of massive irreps. This relationship is discussed in more detail in \cite{Khan:2004nj,Brink:2002zx}.}, 
but we begin in the angle basis where transformation rules are simplest. Eigenstates of $T_{1,2}$,
\be
T_{1,2}  \ket{k,\vec t} = \vec t   \ket{k,\vec t},
\ee
are ``plane-wave'' states in the $\mathbb{R}^2$ on which the $ISO(2)$ Little Group acts.   The states with 
$\vec t_\phi = (\rho \cos\phi, \rho \sin\phi)$ for fixed $\rho$ all have  $W^2 = - \rho^2$ from \eqref{w2massless}, and rotate into one 
another under LG rotations.  Thus states within a single representation of fixed $\rho$ are labeled simply by the polar angle $\phi$, with a periodic identification 
$\ket{2\pi} = \ket{0}$.  
General LG elements \eqref{LGelem} are defined to act on angle-basis states as 
\bea
W[\theta, \beta] \ket{k,\phi} &= &e^{i \vec b.\vec
  t_{\phi+\theta}} \ket{k, \phi+\theta} \label{phiTransform}\\
  &=& e^{i \rho \Re[\sqrt{2}\beta e^{-i(\phi+\theta)}]} \ket{k, \phi+\theta}\\
  &=& \int \frac{d\phi'}{2\pi} D_{\phi\phi'}[\theta,\beta]\ket{k,\phi'}, \quad \mbox{ with } D_{\phi\phi'}[\theta,\beta] = (2\pi) \delta(\phi'-\phi-\theta) e^{i \rho \Re[\sqrt{2}\beta e^{-i\phi'}]}, \label{Dphiphip}
\eea
where we have introduced a two-vector $\vec b= \sqrt{2} ({\rm Re}[\beta],{\rm Im}[\beta])$ in the first line to exhibit the $ISO(2)$ structure, 
and in the last line we have written the transformation rule in the form of \eqref{LGAction}. 
This transformation rule follows from the group structure of ISO(2) (see \cite{Tung:1985na} for example), 
and is therefore consistent with the commutation relations \eqref{WWcommutator}.
Conjugate ``outgoing'' states $\bra{k,\phi}$ transform according to
\bea
\bra{k,\phi} W^{\dagger}[\theta, \vec b] &=& e^{-i \vec b.\vec t_{\phi+\theta}} \bra{k,\phi} \nonumber \\
&=& \int \frac{d\phi'}{2\pi} (D_{\phi\phi'}[\theta,\beta])^* \bra{k,\phi'}.
\eea
The above transformations are unitary with respect to the inner product
\be
\braket{k,\phi}{k',\phi'}= 2k^0\delta^{3}(k-k')\; 2\pi \delta(\phi-\phi') 
\ee
and the Lorentz-invariant sum over states is 
$%\be
\int \frac{d^3\vec{k}}{k^0}\frac{d\phi}{2\pi}.
$%\ee
General little-group states $\ket{\psi_f} = \int \frac{d^\phi}{2\pi} f(\phi) \ket{\phi}$ correspond to arbitrary periodic functions on the circle.

Because the origin $\phi=0$ has no invariant significance, it is natural to associate the angle $\phi$ with a \emph{direction} in space-time, transforming in the natural way\footnote{We thank E. Witten and N. Arkani-Hamed for emphasizing the usefulness of this perspective.}.  
A first step in this direction is to introduce the spacelike unit vector
\be
\epsilon(k,\phi)= \frac{i}{\sqrt{2}}( \epsilon_{+}e^{-i\phi}-\epsilon_{-}e^{i\phi}) = -\sqrt{2} \Im[\epsilon_+ \, e^{-i\phi}] 
\ee
in the plane of $\epsilon_+$ and $\epsilon_-$.  
The transformation of states  \eqref{phiTransform} can be written using \eqref{eq:ALGonEpsilon}--\eqref{eq:BLGonEpsilon} as
\be
W[\theta,\beta] \ket{k,\epsilon(k,\phi)} \simeq e^{i\rho \Re[\sqrt{2} \beta e^{-i(\phi+\theta)}]}  \ket{k,\epsilon'}\quad \epsilon'^\mu \equiv W^\mu_{\nu}\epsilon^\nu -  \Re[\sqrt{2} \beta e^{-i(\phi+\theta)}] k^\mu.
\ee
If we further define an equivalence relation 
\be
\ket{k,\epsilon}\simeq e^{-i\rho a} \ket{k,\epsilon+a k}, \label{epsilonIdent}
\ee
for all real $a$ and with $\epsilon$ constrained only by the covariant requirement $\epsilon.k =0$, the above transformation can be written quite simply as $W \ket{k,\epsilon^\mu} =  \ket{k,W \epsilon}$ and, for general Lorentz transformations $\Lambda$, 
\be
U(\Lambda) \ket{k,\epsilon^\mu} =   \ket{\Lambda k,\Lambda \epsilon}.
\ee
In this formulation, checking the Little Group covariance of amplitudes and of the soft factor \eqref{CSRsoft} is quite simple: when written in terms of $\epsilon(k,\phi)$, they must be simultaneously invariant under the identification \eqref{epsilonIdent} and under Lorentz transformations where $\epsilon$ is taken to transform covariantly.

The \emph{spin basis}, obtained by Fourier transforming in $\phi$, diagonalizes $R$ and makes contact with the helicity representations.
Writing
\be
\ket{k,n} \equiv \int \frac{d\phi}{2\pi} e^{in\phi} \ket{k,\phi} 
\ee
for integer $n$, we obtain the transformation rule
\bea
W[\theta, \beta] \ket{k,n} & = & \sum_{n'} D_{n n'}[\theta,\beta] \ket{k,n'} , \label{eq:spintransformations}\\
D_{n n'}[\theta,\beta] & = & \int \frac{d\phi d\phi'}{(2\pi)^2} \ D_{\phi\phi'}[\theta,\beta] e^{i n\phi}e^{-i n' \phi'} \nonumber \\
&=& e^{-in\theta}(ie^{i\alpha})^{(n-n')}J_{n-n'}(\rho\sqrt{2}|\beta|) \label{eq:Ddefinition},
\eea
where $\beta=|\beta|e^{i\alpha}$.
Outgoing states transform as
\be
\bra{k,n}W^{\dagger}[\theta, \beta] = \sum_{n'} \bra{k,n'} (D_{n n'}[\theta,\beta])^*.
\ee
The appearance of Bessel functions is to be expected, as they are representation functions for the Euclidean 
group in two dimensions (see for example \cite{Tung:1985na} or \cite{VilenkinKlimyk}). 
These transformations are unitary with respect to the inner product 
\be
\braket{k',n'}{k,n}= \delta_{n' n} 2k^0\delta^{3}(k-k').
\ee
This result and many other relations in the spin basis follow from the Bessel function addition theorem\footnote{Namely $e^{i n\theta}J_n(R)=\sum_k e^{i k \phi}J_k(r)J_{n-k}(r')$ for $r+r'e^{i\phi}=Re^{i\theta}$ -- see e.g. \cite{Tung:1985na}.}, 
which is a simple consequence of $ISO(2)$ representation theory. 

Finally we define the full unitary action of Lorentz transformations on single-particle CSP states in the usual way,
\bea
U(\Lambda)\ket{k,\phi} &=& W_{\Lambda, k}\ket{\Lambda k,\phi} \\
&=& \int \frac{d\phi'}{2\pi} D_{\phi\phi'}[\theta_{\Lambda,k},\beta_{\Lambda,k}]\ket{\Lambda k,\phi'}, \\
U(\Lambda)\ket{k,n} &=& \sum_{n'}D_{n n'}[\theta_{\Lambda,k},\beta_{\Lambda,k}]\ket{\Lambda k,n'},
\eea
with $W[\theta_{\Lambda,k},\beta_{\Lambda,k}] \equiv W_{\Lambda,k} = B_{\Lambda k}^{-1} \Lambda B_k$ used to define the Little Group rotation 
$\theta_{\Lambda,k}$ and translation $\beta_{\Lambda,k}$ induced by $\Lambda$. With the standard boost $B_k$ specified by a choice of frames
for all $k$, the above unitary action unambiguously defines the action of Lorentz transformations on single-particle states.   

We note that, for an appropriate choice of $\epsilon_\pm$ (specifically $\epsilon_{\pm}^0 = 0$), $n$ is precisely the eigenvalue of the 
``helicity'' operator \eqref{helicity3V}, but for $\rho\neq 0$ it acts more like a ``spin'' label than a helicity label, in that it is not boost-invariant.  
In the limit $\rho\rightarrow 0$, however, $J_{n-n'}(\rho|2\beta|)$ approaches zero for $n\neq n'$ and $1$ for $n=n'$, so the transformation rule reduces to 
\be
D_{n n'}[\theta,\beta] \rightarrow e^{-in\theta}\delta_{n n'}.
\ee
In other words, we recover in the $\rho\rightarrow 0$ limit a direct sum of all integer-helicity states.  

A second ``double-valued'' type of continuous-spin representation for each $\rho$ is obtained by making the antiperiodic identification $\ket{2\pi}=-\ket{0}$ in \eqref{phiTransform} instead of the periodic identification.  In this case, the spin-basis states are labeled by half-integer $n$.  As these are distinct representations, their states do not mix.  We will focus throughout this paper on the single-valued case.
For further discussion of double-valued continuous-spin representations, CSPs in higher dimensions, and supersymmetrized CSPs, we refer the reader to 
\cite{Brink:2002zx, Khan:2004nj, Mourad:2005rt,Bekaert:2005in, Edgren:2005gq, Zoller:1991hs}.

\subsection{Parity and Time-Reversal Transformations}
It is also possible to define actions of parity and time-reversal on single-CSP states, which will be useful when we discuss crossing symmetry. 
Following the conventions of Weinberg \cite{WeinbergQFT} we introduce a linear and unitary parity operator $P$ and an antilinear and antiunitary 
time-reversal operator $T$ consistent with Poincare invariance. 
This requires that 
\begin{alignat}{2}
PHP^{-1} &= H & \qquad THT^{-1} &= H \\
P\vec{P}P^{-1} &= -\vec{P}& \qquad T\vec{P}T^{-1} &= -\vec{P}\\
P\vec{J}P^{-1} &= \vec{J}& \qquad  T\vec{J}T^{-1} &= -\vec{J} \\
P\vec{K}P^{-1} &= -\vec{K}& \qquad T\vec{K}T^{-1} &= \vec{K}  
\end{alignat}
where $J_{ij}=\epsilon_{ijk}\vec{J}_k$ and $J_{0i}=\vec{K}_i$ are the rotation and boost generators, respectively. 

A straightforward calculation (following \cite{WeinbergQFT}) shows that the only consistent action of $P$ and $T$ on 
spin basis states (for integer spins) is
\bea
P\ket{p,n} &=& \eta \ket{\bar p, -n} \\
T\ket{p,n} &=& \xi (-1)^n\ket{\bar p, n},
\eea
with $\eta$ and $\xi$ arbitrary (but $n$-independent) phases. In the angle basis, 
\bea
P\ket{p,\phi} &=& \eta \ket{\bar p,-\phi} \\
T\ket{p,\phi} &=& \xi \ket{\bar p, -\phi+\pi}
\eea
so that $PT$ acts simply as
\bea
PT \ket{p,n} &=& \eta\xi (-1)^n\ket{p,-n} \\
PT \ket{p,\phi} &=& \eta\xi \ket{p,\phi+\pi}. \label{eq:PT}
\eea
The counter-intuitive lack of a sign flip in $\phi$ under $PT$ follows from our use of an antilinear $T$. 

Throughout this paper, we will not consider CSPs with additional quantum numbers that would require the existence of 
distinct anti-particles. For that reason, we assume that charge conjugation maps
CSP states to themselves. 

%%%%%%%%%%%%%%%%%
\section{Wavefunctions}\label{sec:wavefunctions}
This section reviews the notion of an ``auxiliary-space'' wavefunction $\psi(\eta,x)$, identifies the most general such wavefunction for a CSP, and uses these wavefunctions to find new wave equations for CSPs.  It is surprising that this was not done long ago --- almost all studies of CSPs in the last seven decades (excepting \cite{Iverson:1971hq}) have assumed Wigner's CSP wave equations \cite{Wigner:1947,Bargmann:1948ck}, although Wigner himself noted that more than one wave equation may describe the same type of particle.  

The above logic --- constructing wavefunctions that behave like single-particle states under Lorentz transformations, then identifying the covariant equations they solve --- is the reverse of the modern approach where wave equations are derived from a covariant action, and in turn used to identify a basis of particle-like solutions.  Yet the bottom-up approach, pioneered by Majorana, Dirac, and others in the 1930's and formalized by Wigner and Bargmann, has proved extremely useful historically \cite{Majorana, Dirac:1936tg,Proca, Fierz:1939ix,Rarita:1941mf,Wigner:1947,Bargmann:1948ck} (see also \cite{Sorokin:2004ie} and \cite{WeinbergQFT} ch. 5 for modern treatments).  

A covariant wavefunction in this sense is an object with Little Group \emph{and} Lorentz labels ($a$ and $l$ respectively), whose defining property relates the action of a generic Lorentz transformation $\Lambda$ to the unitary (but momentum-dependent) Little Group action it induces:
\be
\sum_{a'} D_{a a'}\left[W_{\Lambda,k}\right] \psi\left(\mathbf{\Lambda k},a', l \right) 
= \sum_{\bar l} D^{-1}_{l \bar l}[\Lambda] \psi\left(\mathbf{k},a, \bar l \right),
\label{wfnCovariance}
\ee
where 
$D_{l \bar l}$ and $D_{a a'}$ are Lorentz and Little Group representation 
matrices, and $W_{\Lambda,k}$ is the Little Group transformation induced by $\Lambda$, as defined in \eqref{particleDef}
\footnote{An important caveat to this approach is that the covariant wave equations for gauge fields do not have solutions that 
are covariant in the sense of \eqref{wfnCovariance} --- rather, these are covariant \emph{up to a gauge transformation} that 
leaves the equations of motion invariant.  We have not classified solutions of this type, though we have found evidence that 
gauge theories of CSPs are particularly useful \cite{Schuster:2013pta}.}.
This is precisely the condition satisfied by coefficient functions used in building covariant fields from creation and annihilation operators, though we will not use our wavefunctions for that purpose.  It is also reminiscent of the $S$-matrix \eqref{COVARIANCE} and soft factor \eqref{softCovariance} covariance equations.  

We find two classes of new wavefunctions that generalize solutions to Wigner's equations and the wavefunctions of \cite{Iverson:1971hq} respectively, and each class suggests a family of new wave equations.  The wavefunctions of the latter class satisfy two important properties: the wavefunctions are ``smooth'' in a sense that facilitates using them to build soft factors, and the associated wavefunctions reduce at $\rho=0$ to a gauge-fixed form of the Fronsdal helicity-$h$ equations of motion. 
 
\subsection{Auxilliary-Space Wavefunctions}
Because the continuous-spin representation of the \Poincare Group is infinite-dimensional, we will need to build wavefunctions that also transform in infinite-dimensional Lorentz representations.  
The general \emph{irreducible} representations of $\SLtC$, the covering group of the Lorentz group, were classified by Gelfand \cite{Gelfand}.  
Whereas objects with $\SLtC$ spinor indices (e.g. $\psi^a$ or $\psi^{ab\dot a\dot b}$) transform in finite-dimensional representations, the infinite-dimensional representations naturally act on homogeneous functions $\psi(\xi^a,\bar\xi^{\dot a})$ of a complex spinor $\xi^a$.  

It is often useful, however, to work with wavefunctions that transform in \emph{reducible} representations of the Lorentz group.  For example, we usually describe gravitons using a tensor field $g^{\mu\nu}$ whose trace is unconstrained.  In a similar spirit, we will consider wavefunctions $\psi(\eta)$ in the space of functions of an auxiliary vector $\eta^\mu$ --- an infinite-dimensional (reducible) representation on which Lorentz transformations act as
\be
D[\Lambda] \psi(\eta, x) \equiv \psi(\Lambda \eta, \Lambda x) \label{eq:Ltr}.
\ee

We note that this Lorentz action leaves invariant the subspaces of rank-$n$ polynomials in $\eta$ (e.g. $\psi(\eta) = h^{\mu\nu} \eta_\mu \eta_\nu$).  On these spaces the transformation \eqref{eq:Ltr} is equivalent to the usual transformation of the symmetric tensor coefficient, 
\be
D[\Lambda] h^{\mu\nu} = {\Lambda^\mu}_{\mu'}  {\Lambda^\nu}_{\nu'} h^{\mu' \nu'}.
\ee
In this way, the homogeneity-$n$ functions of $\eta$ may be regarded as an infinite-dimensional extension of the rank-$n$ symmetric tensor representations. 

\subsection{Covariant Wavefunctions and Wave Equations}
In the notation developed above, the covariance equation \eqref{wfnCovariance} for CSPs in the $\phi$ basis is simply
\be
\int \frac{d\phi'}{2\pi} D_{\phi \phi'}\left[W(\Lambda,k)\right] \psi \left( \{ \mathbf{\Lambda k},\phi' \},\eta^\mu \right) 
= \psi \left( \{ \mathbf{k},\phi \}, \Lambda^{-1} \eta \right). \label{covariancePhi}
\ee
For the special case $\Lambda = B_{k'} B_{k}^{-1}$, we have $W(\Lambda,k) = {\mathbf 1}$ so that 
\be
\psi \left( \{ \mathbf{\Lambda k},\phi \},\Lambda \eta \right)
= \psi \left( \{ \mathbf{k},\phi \}, \eta \right). \label{Bkcondition}
\ee
This is satisfied if and only if $\psi$ is a ``scalar''-valued function of $k$, $\eta$, and $\epsilon_{\pm}(k)$ (we recall that $\epsilon_\pm(k)^\mu$ transform as tensors under the $B_k$'s, so contractions $\epsilon_\pm(k).\eta$ satisfy \eqref{Bkcondition} even though they are not true scalars under general Lorentz transformations). Because any Lorentz transformation can be decomposed into a product of $B_k$'s and Little Group elements, any ``scalar'' $\psi$ that solves \eqref{covariancePhi} for Lorentz transformations $W\in LG_k$ will also solve \eqref{covariancePhi} for general $\Lambda$.  

Taking $W$ to be an infinitesimal Little Group transformation $W = 1+ \frac{i}{\sqrt{2}}\beta T_{-} + \frac{i}{\sqrt{2}}\beta^*T_{+}-i \theta R$, we can use \eqref{LGgenAction} to convert \eqref{covariancePhi} into a system of differential equations:
\bea
-i\left( \eta.\epsilon_{-} \epsilon_{+}.\partial_{\eta}-\eta.\epsilon_{+} \epsilon_{-}.\partial_{\eta} \right) \psi &=& \partial_{\phi} \psi \\ \label{Reom}
-\left( \eta.\epsilon_{-}p.\partial_{\eta} - \eta.p\epsilon_{-}.\partial_{\eta} \right) \psi &=& \frac{\rho}{\sqrt{2}} e^{-i\phi} \psi \\ \label{T-eom}
\left( \eta.\epsilon_{+}p.\partial_{\eta} - \eta.p\epsilon_{+}.\partial_{\eta} \right) \psi  &=& \frac{\rho}{\sqrt{2}} e^{i\phi} \psi \label{T+eom}.
\eea
These equations are homogeneous in $\eta$ and Fourier-conjugate to themselves.
Any family of solutions $\psi(\{p,\phi \},\eta)$  to this system of equations forms a basis of solutions to a particular covariant wave equation.
Not surprisingly, these equations imply
\be
\left( W^2 + \rho^2 \right) \psi = 0,
\ee
where $W^2$ can be computed from the Lorentz action on $\psi$ (also assuming $p^2=0$) as
%mostly negative metric form of W^2
\be
W^2 = 2p.\eta \, p.\partial_{\eta}\, \eta.\partial_{\eta} - (p.\eta)^2\partial_{\eta}^2 - \eta^2(p.\partial_{\eta})^2.
%+ p^2 \eta^2 \partial_{\eta}^2 -p^2 \eta\cdot\partial_{\eta}(\eta\cdot\partial_{\eta}+1)
\ee
Here, the Pauli-Lubanski $W^{\mu}$ is constructed using Lorentz generators $J_{\mu\nu}$ that 
act on wavefunctions according to \eqref{eq:Ltr}. Thus, $J^{\mu\nu}\propto (\eta^{[\mu}\partial_{\eta}^{\nu]}+p^{[\mu}\partial_{p}^{\nu]})$,
only the first term contributes to $W^\mu$, and $W^2$ can be explicitly calculated yielding the above. 

The derivation of the full set of solutions is instructively described in Appendix B; here we summarize only our key results. 
There are two classes of solutions -- those that are smooth in $\eta$ near $\eta.p$=0, and singular solutions supported on $\delta(\eta.p)$.
Importantly, general wavefunctions for CSPs {\bf need not} satisfy Wigner's original wave equations!
While the singular solutions are related to a basis of solutions to the Wigner equations, the smooth solutions solve a 
new class of wave equations, described below, that make contact with the Fronsdal equations \cite{Fronsdal:1978rb} when $\rho=0$. 

The \textbf{singular solutions} are
\be
\psi(\{p,\phi,f\},\eta)=\int dr f(r) \int d\tau \delta^{4}(\eta-r\epsilon(p\phi)-r \tau p)e^{-i\tau\rho}, \label{planewave}
\ee
where $f(r)$ specifies an arbitrary profile of $\psi$ under re-scalings of $\eta$. For generic $f(r)$, 
$\eta$ has support on a plane spanned by $\epsilon(p\phi)$ and $p$, with
\be
\epsilon(p\phi) \equiv \frac{i}{\sqrt{2}}( \epsilon_{+}e^{-i\phi}-\epsilon_{-}e^{i\phi}) = -\sqrt{2} \Im[\epsilon_+ \, e^{-i\phi}]\label{eq:Epsilon}.
\ee
These solutions satisfy the Lorentz-covariant wave equations
\bea
p^2\psi &=& 0 \label{e1sing}\\ 
p\cdot\eta \psi &=& 0 \\
\left( -\eta^2(p\cdot\partial_{\eta})^2+\rho^2 \right) \psi &=& 0, \label{e3sing}
\eea
where the last equation is simply $W^2+\rho^2=0$ on the support of the other two equations of motion.  
These three equations, together with $\left( \eta^2+1 \right) \psi =0$, are the Wigner equations for continuous-spin particles.  
To recover a basis of solutions to the Wigner equation we may choose $f(r)=\delta(r-1)$, but we stress that this choice is not unique.  
Any $p$-independent equation of motion will single out a particular function $f(r)$, and the resulting $\psi(\{p,\phi,f\},\eta)$ constitute a basis of covariant wavefunctions satisfying the chosen $\eta$-space equation and the system \eqref{e1sing}-\eqref{e3sing}.  
For example, \eqref{planewave} with $f(r)=r^3$ satisfy \eqref{e1sing}-\eqref{e3sing} and the homogeneity condition 
$\eta\cdot\partial_{\eta}\psi=0.$

The \textbf{smooth solutions} are
\bea
\psi(\{p,\phi\},\eta) &=& f(\eta.p,\eta^2) e^{i\rho \frac{\eta.\epsilon(p\phi)}{\eta.p}} \nonumber \\
&=&   f(\eta.p,\eta^2) e^{-i\rho\sqrt{2} \Im \left( \frac{\eta.\epsilon_{+}e^{-i\phi}}{\eta.p} \right) }, \label{smoothWavefunction}
\eea
where $f(\eta.p,\eta^2)$ is another arbitrary function.  
These solutions, involving an arbitrary function of two variables, can satisfy an even broader array of Lorentz-covariant wave equations
than the singular solutions. 
All of the smooth solutions satisfy 
\bea 
p^2 \psi & = & 0 \\
(W^2 + \rho^2)\psi & = & 0. 
\eea
Two more wave equations can be chosen arbitrarily to fix the functional form of $f(\eta.p,\eta^2)$.
One convenient choice is 
\bea
p\cdot\partial_{\eta} \psi &=& 0 \label{eq:eomfirst} \\
\eta\cdot\partial_{\eta} \psi &=& n \psi
\eea
for any integer $n$, on which the $W^2$ equation simplifies to 
\bea
p^2 \psi &=& 0 \\
\left( (p\cdot\eta)^2\partial_{\eta}^2 + \rho^2 \right) \psi &=& 0 \label{eq:eomlast}
\eea
In particular, the choice $n=0$ implies $f=1$.
The equations of motion \eqref{eq:eomfirst}-\eqref{eq:eomlast} are new. 
Importantly, for any non-negative $n$, in the limit $\rho = 0$, we recover the transverse/traceless gauge-fixed form of
 the Fronsdal equations for helicity-$n$ particles \cite{Fronsdal:1978rb}. This connection will be interpreted and exploited 
in \cite{Schuster:2013pta}.

In summary, we have exhibited new equations of motion for CSPs that make direct contact when $\rho= 0$ with familiar equations of motion helicity particles. In the next section, we will use these new objects 
to build suitably smooth scattering amplitudes.

%%%%%%%%%%%%%%%%%%%%%%%%%%%%%%%%%%%%%%%%
\section{Lorentz-Invariant S-Matrix for Soft CSP Emission}\label{sec:WFandSoftFactor}
Together, Lorentz-invariance and unitarity impose significant constraints on how different massless particles can interact. These constraints 
are particularly simple in the case of amplitudes for emission of a single, low-energy massless particle.  Famously, Weinberg found these soft 
limits to be so restrictive that one can derive the most salient features of helicity-$h$ interactions ---
charge-conservation (the equivalence principle) in helicity-1 (2) coupling,  and the impossibility of 
$h>2$ couplings strong enough to mediate a long-range force --- from soft limits alone \cite{Weinberg:1964ew}.  (Similar arguments apply in formalisms where Lorentz-covariance is manifest but unitarity is not~\cite{Benincasa:2007xk,Schuster:2008nh})
Because CSPs resemble an infinite 
tower of high-helicity modes, one might expect CSP soft-emission amplitudes to have problems similar to those of high-helicity particles.  
We will show, however, that no such problems arise --- a rather striking result in which the boost non-invariance of CSP spins plays a crucial role!  
Our framing of the problem largely follows \cite{Weinberg:1964ew}.  

\subsection{Lorentz-Invariance and Unitarity Constraints on Scattering Amplitudes}
We assume that scattering amplitudes for CSPs are defined by matrix elements of a unitary $S$-matrix 
\be
A({p_i, a_i}\rightarrow {p_j,a_j}) = \bra{{p_j, a_j}} S \ket{{p_i,a_i}},
\ee
where $p_i$ and $a_i$ ($p_j$ and $a_i)$ are the momenta and LG labels of initial (final) states.
Poincar\'e-invariance of the $S$-matrix, $[S,U(\Lambda)]=0$, implies that all scattering amplitudes satisfy the covariance equation
\be
A(\{p_i, a_i\}\rightarrow \{p_j,a_j\}) =\left( \prod_j \sum_{a'_j} (D_{a_j a'_j}[W_{\Lambda, p_j}])^* \right)\left( \prod_i \sum_{a'_i} D_{a_i a'_i}[W_{\Lambda, p_i}] \right)
A(\{\Lambda p_i, a'_i\}\rightarrow \{\Lambda p_j,a'_j\}). \label{COVARIANCE}
\ee
The reader will recognize a close resemblance of this equation to the wavefunction covariance equation \eqref{wfnCovariance}: scattering 
amplitudes transform under the Little Group as a product of single-particle states, and trivially under the Lorentz group.  
This suggests that a natural way of forming Lorentz-covariant amplitudes is to contract single-particle wavefunctions' Lorentz indices with one 
another and with functions of momenta. 

But what is the analogue of index contraction for an auxiliary-space wavefunction $\psi(\eta,\{k,\phi\})$?  This has no indices, but its Lorentz-transformation is encoded in the auxiliary spinor $\eta$, which should drop out of the final answer.  We can ``contract'' $\eta$ by integrating over it: 
\be
\int d^4 \eta f(\eta,p_i, \dots) \psi(\eta, k),\label{integralContract}
\ee
The simplest example, which we will return to, is $\psi(\eta=p_i,k)$ for some particles' momentum $p_i$.  In the special case of a polynomial wavefunction $\psi(\eta,x)=A^{\mu}(x)\eta_\mu$, this is of course equivalent to the usual index contraction.

Building amplitudes from covariant wavefunctions is \emph{sufficient} to guarantee covariance of the amplitudes,  but not necessary.  Indeed, standard helicity-1 and 2 amplitudes are \emph{not} of this form, when written in a local and manifestly unitary form.  
Rather, these amplitudes are conventionally built by contracting wavefunctions $\prod_{i=1}^h\epsilon_\pm^{\mu_i}({\bf k})$, which are only 
Little-Group covariant \emph{up to a gauge term proportional to $k^\mu$}, into a symmetric function ${\cal M}_{\mu_1 \dots \mu_h}$ that 
satisfies $k_{\mu_1} {\cal M}_{{\mu_1} \dots {\mu_h}} = 0$.  Indeed, it has been shown using $S$-matrix properties alone that any 
helicity-$h$ amplitude can be expressed in this form (see \cite{WeinbergQFT} and references therein).  Likewise, we suspect that more 
general ``contractions'' of non-covariant CSP wavefunctions may be useful in building some amplitudes, even though we will not need them to construct single-emission amplitudes.  

\begin{figure}[!htbp]
\includegraphics[width=0.9\columnwidth]{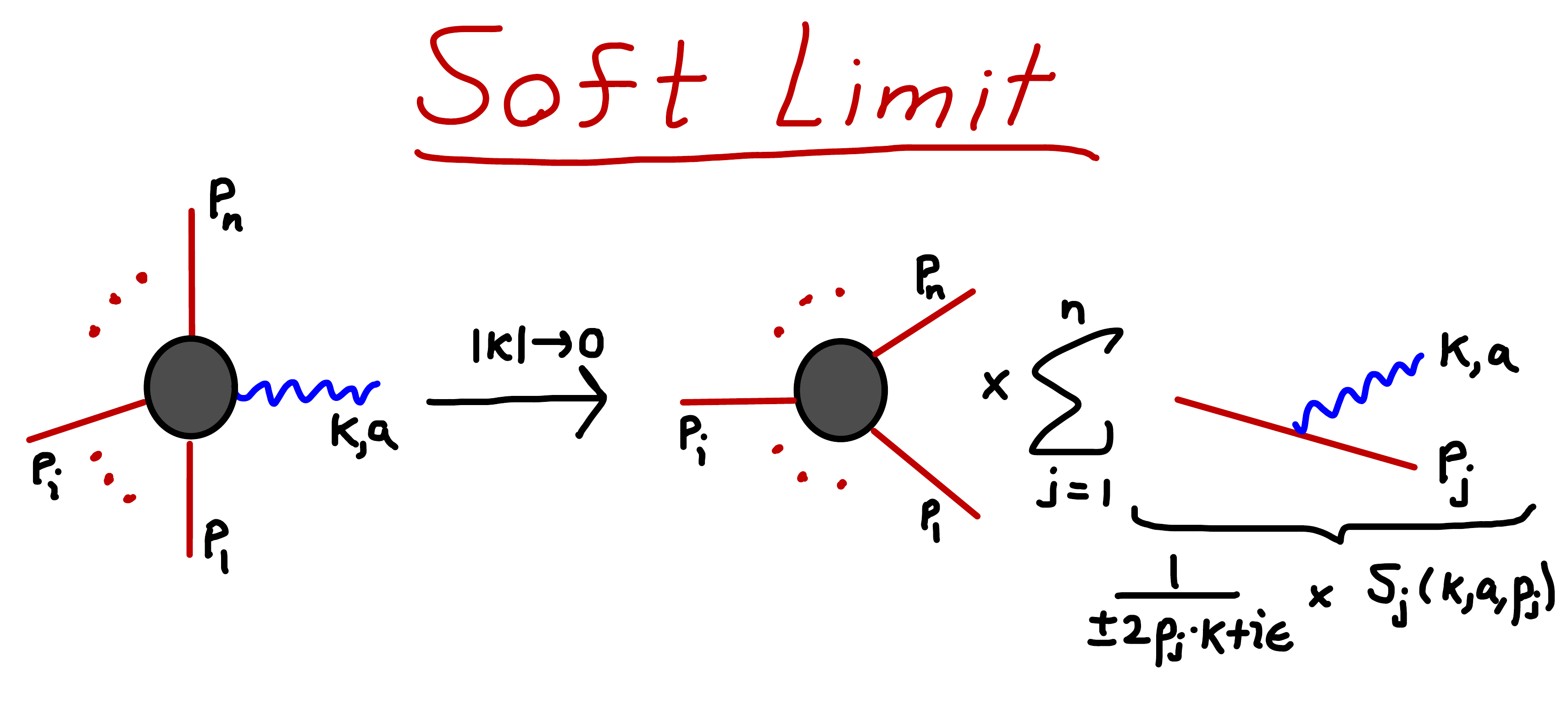}
\caption{A diagrammatic picture for the factorization of single-emission amplitudes in the limit that the leg $\{k,a\}$ becomes soft.
In this limit, radiation of $k$ off external legs grows as $|{\bf k}|\rightarrow 0$ and therefore dominates the amplitude.  Each single emission term is proportional to a ``parent'' amplitude times a ``soft factor'' depending only on $k$, $a$, and a single external momentum $p_i$.  The sum of these single-emission terms must be Lorentz-invariant up to terms of order $|{\bf k}|^0$. 
\label{fig:soft}}
\end{figure}

We turn now to unitarity, whose simplest consequence is the tree-level pole structure of scattering amplitudes
\footnote{Like Weinberg's classic paper, we neglect subtleties associated with infrared modifications to these poles.}  
This pole structure implies a particularly simple form for amplitudes involving $n$ particles of momentum $p_1,\dots, p_n$, plus a 
single massless ``soft particle'' whose 
momentum $k$ satisfies  $k.p_i \ll p_i.p_j$ for all $i,j$.  As $k$ gets smaller, one kind of contribution will grow, and eventually dominate: the 
diagrams where the soft particle is emitted from one of the external legs (see Figure \ref{fig:soft}).  These diagrams have a propagator that 
grows without bound as $k$ gets softer:  
\be
\frac{1}{(p_i\pm k)^2 - m_i^2 + i\epsilon} = \frac{1}{(p_i^2-m^2) \pm 2 p_i.k + i\epsilon}  = \frac{1}{\pm 2 p_i.k + i \epsilon},
\ee
where the top sign refers to outgoing particles, and the bottom to incoming particles (a convention we maintain throughout this section).  
In contrast, propagators from ``internal'' emissions will be suppressed by the scale $p_i.p_j$.  Unitarity further implies that these ``external emission'' 
contributions to the amplitude factorize on the pole as 
\be
A_0(p_1,\dots,p_n) \times \frac{1}{\pm 2 p_i.k + i\epsilon} \times s_i(\{k, a\},p_i)_- + {\cal O}(|k|^0), \label{softFactorization}
\ee
where $A_0$ is a ``parent amplitude'' involving the $n$ hard particles but not the soft particle, $s_i$ is a ``soft factor'', and $a$ denotes 
the Little Group state of the soft particle.  We adopt the conventions $s_i(\dots)_-$ for emission soft factors, and $s_i(\dots)_+ = s_i(\dots)_-^*$ for absorption soft factors.

As external emissions dominate at small $k$, they must separately satisfy \eqref{COVARIANCE}, at least up to corrections that 
are not singular as $|{\bf k}|\rightarrow 0$.  Since $A_0$ is itself an amplitude of the $n$ hard particles that satisfies 
\eqref{COVARIANCE}, applying \eqref{COVARIANCE}  to the full amplitude leads to the constraint
\bea
f(\{k, a\},p_1,\dots,p_n)_- & = & \sum_{a'} D_{a a'}^*[W_{\Lambda, k}] f(\{\Lambda k, a'\},\Lambda p_1,\dots,\Lambda p_n) + {\cal O}(|k|^0),\label{softCovariance}\\
\mbox{where } f(\{k, a\},p_1,\dots,p_n)_- & \equiv & \sum_{i=1}^n \frac{1}{\pm 2 p_i.k + i\epsilon} \times s_i(\{k, a\},p_i)_-.
% = \sum_{i=1}^n \frac{1}{\pm 2 p_i.k + i\epsilon} \times \sum_{a'} D_{a' a}^{-1}[W_{\Lambda, k}]  s_i(\{\Lambda k, a'\},\Lambda p_i)
\label{softUnitarity}
\eea
A sufficient condition (but not necessary) is covariance of the soft factors themselves, i.e. 
\be
s_i(\{k, a\},p_i)_-  =  \sum_{a'} D_{a a'}^{*}[W_{\Lambda, k}] s_i(\{\Lambda k, a'\},\Lambda p_i)_-,\label{softCovariance1}
\ee
which is very similar indeed to the wavefunction covariance equation \eqref{covariancePhi}.

At this point, it is quite easy to see how Weinberg's constraints on high-helicity interactions arise. For helicities higher than 2, 
there is simply no function $f$ that simultaneously satisfies the covariance requirement \eqref{softCovariance} and the 
decomposition \eqref{softUnitarity} demanded by unitarity!  Using the decomposition of helicity amplitudes mentioned above, 
\eqref{softCovariance} is equivalent to the requirement
\bea
f(\{k, \pm h\},p_1,\dots,p_n)_- = \epsilon_\pm^{*\mu_1}({\bf k}) \dots \epsilon_\pm^{*\mu_h}({\bf k}) {\cal M}_{\mu_1 \dots \mu_h}(p_1,\dots, p_n, k),\\
\mbox{ with }  k_{\mu_i}{\cal M}_{\mu_1 \dots \mu_h}(p_1,\dots, p_n, k) = 0.
\eea
Since each $s_i$ can only depend on $p_i$, and $\{k,a\}$, unitarity implies a unique form
\be
{\cal M}_{\mu_1 \dots \mu_h}(p_1,\dots, p_n, k) = \sum_i \frac{1}{\pm 2 p_i.k + i\epsilon} \times g_i {p_i}_{\mu_1} \dots {p_i}_{\mu_h} + {\cal O}(|k|^0)
\ee
and hence Lorentz-invariance requires 
\be
0=k_{\mu_1} {\cal M}_{\mu_1 \dots \mu_h}(p_1,\dots, p_n, k) = \sum_i \f{\pm 1}{2} g_i {p_i}_{\mu_2} \dots {p_i}_{\mu_h} + {\cal O}(|k|^1),\label{WeinbergFinal}
\ee
where we have cancelled a numerator factor in each term of $k.p_i$ from the contraction of $k$ into ${\cal M}$ against the propagator. 
For $h=\pm 1$, the condition is simply $\sum_i \pm g_i = 0$ --- charge conservation.  For $h=\pm 2$, it is linear in $p_i$, and can 
only be solved at generic momenta if $g_i = g$ is universal (a form of the equivalence principle), in which case it reduces to momentum
 conservation.  For higher helicities, the equation is quadratic or higher-order in $p_i$, and therefore no choice of $g_i$ will 
 satisfy \eqref{WeinbergFinal} for generic momenta $\{p_i\}$! The crux of these obstructions was that the amplitudes were built out of 
 wavefunctions (products of $\epsilon$'s) that were \emph{almost, but not quite} covariant.  

\subsection{Covariant Soft Factors for Continuous-Spin Particles}
We are now ready to describe our soft factor ansatz for CSPs, obtained from the smooth wavefunctions 
\eqref{smoothWavefunction} by evaluating them at $\eta^\mu = p_i^\mu$ for each $p_i$.  In the angle basis, we can take  
\bea
s_i(\{k,\phi\},p_i)_{\pm} &= & f_i^{\pm}(k.p_i, m_i^2) e^{\pm i \rho \frac{\epsilon(k\phi).p_i}{k.p_i}}\\
& = & f_i^{\pm}(k.p_i, m_i^2) e^{\mp i \rho \sqrt{2} \Im\left[e^{-i\phi} \frac{{\epsilon^+}.p_i}{k.p_i}\right]}\label{CSRsoft},
\eea
for any function $f^+={f^-}^*$. 
We obtain soft factors in the spin basis by Fourier-transform, finding 
\bea
s_i(\{k,n\},p_i)_{\pm}  & = & \int \frac{d\phi}{2\pi} e^{\pm i n\phi} s_i(\{k,\phi\},p_i)_{\pm} \\
& = & f_i^\pm(k.p_i, m_i^2) (-1)^n  e^{\pm i n \arg\left[\frac{\epsilon_{+}\cdot p_i}{k\cdot p_i}\right]} J_n\bigg(\rho \sqrt{2}\bigg|\frac{p.\epsilon_{+}}{k.p}\bigg|\bigg).\label{Soft_spinBasis}
\eea
The behavior of the phase in \eqref{CSRsoft} as $p_i.k \rightarrow 0$ is peculiar, though we do not think it is physically problematic 
(as discussed further in \S\ref{analytic}).  One might also worry that its form invalidates our assumptions that external emissions dominate 
the amplitude in the soft limit --- after all, the Laurent expansion of this phase contains arbitrary negative powers of of $p.k$.  However, 
the phase in \eqref{CSRsoft} has unit magnitude for any real momenta, so external emissions are guaranteed to dominate 
over internal ones in the soft limit provided that $f_i$ does not vanish at $k.p_i=0$.  
 We focus here on the simplest case, a momentum-independent $f_i(k.p_i)=c_i$. Certain soft factors with momentum-dependent $f_i$ also dominate over 
 internal emissions in an appropriate soft limit.  These are precisely the soft factors that exhibit a high-energy correspondence with helicity 1 and 2 soft factors, discussed in \cite{Schuster:2013vpr}.  

Before exploring the physics of these soft factors, let us ask what other general forms one might expect.  We cannot exclude more general soft factors that are not individually covariant, nor more complex covariant functions of multiple particle momenta.  But the simplicity of soft factors (which can depend \emph{only} on $k$ and $p_i$) implies that substituting $p_i\leftrightarrow \eta$ in any covariant soft factor must yield a wavefunction satisfying \eqref{covariancePhi}, which were previously classified into smooth wavefunctions (used to motivate the soft factor \eqref{CSRsoft}) and singular ones.  The replacement $\eta\rightarrow p_i$ in a singular wavefunction would yield a candidate soft factor with singular support in momentum-space, violating cluster decomposition.  
The focus in the literature on singular wavefunctions may be one reason why CSP soft factors were not proposed earlier.  Integrals \eqref{integralContract} of singular wavefunctions against suitably smooth $f(\eta,p_i)$ can have smooth momentum-support, but are always equivalent to \eqref{CSRsoft}.  
This uniqueness underscores the potential importance of the soft factors \eqref{CSRsoft}. 
 
\subsection{Amplitudes from Soft Factors}
To clarify the role of soft factors in CSP scattering amplitudes, 
consider a $2\rightarrow 3$ scattering process involving familiar scalar particles
labeled by $p_1,...p_4$ and an emitted CSP labeled by $k,\phi$. In the limit of $k\rightarrow 0$,
we expect the poles associated with the intermediate scalar particles going on-shell to dominate the 
amplitude, so that 
\be
A(p_1,p_2,p_3,p_4,\{k,\phi \}) \rightarrow A_0(p_1,p_2,p_3,p_4)\sum_{i=1}^4 \frac{g_i}{2p_i\cdot k + i\epsilon} s_i(\{k,\phi \},p_i)_{-}
\ee
where $A_0(p_1,p_2,p_3,p_4)$ is the underlying $2\rightarrow 2$ process, and $g_i$ are couplings associated with 
CSPs attached to scalar particle legs. 
In contrast to the standard helicity soft factors, soft emission amplitudes of this form are Lorentz-covariant term by term. 
Covariance alone is insufficient to derive constraints on the couplings of different legs.

It is instructive to explicitly check that a scattering amplitude of the above form is indeed covariant
in the sense of \eqref{COVARIANCE}.
For an amplitude involving $n$ scalars and one outgoing CSP, Lorentz covariance of the S-matrix \eqref{COVARIANCE} is just
\be
A(\{ k,\phi \}, \{ p_i \}) = \int \frac{d\phi'}{2\pi} (D_{\phi\phi'}[W(\Lambda, k)])^* A(\{ \Lambda k,\phi' \}, \{ \Lambda p_i \}), \label{eq:singleCovariance}
\ee
where $D_{\phi\phi'}$ is given by \eqref{Dphiphip}. 
We need to know how $\epsilon(k,\phi)$ (defined by \eqref{eq:Epsilon}) transforms to check this. 
Recall that $\epsilon^{\mu}_{\pm}(k)\equiv (B_k)^{\mu}_{\nu}\epsilon^{\nu}_{\pm}(\bar k)$, where
$\bar k$ is the reference momentum, and that a general Lorentz transformations $\Lambda$ is decomposed 
in terms of $B_k$ and Little Group transformations as 
$\Lambda \equiv B_{\Lambda k} W[\theta(\Lambda,k),\beta(\Lambda,k)] (B_k)^{-1}$.
Using this, as well as the action of $W[\theta(\Lambda,k),\beta(\Lambda,k)]$ on $\epsilon_{\pm}(\bar k)$
computed using \eqref{eq:ALGonEpsilon}-\eqref{eq:BLGonEpsilon}, we have 
\be
\Lambda^{\mu}_{\nu}\epsilon(k,\phi)^{\nu} = \epsilon(\Lambda k, \phi+\theta)^{\mu} + Re[\sqrt{2} \beta e^{-i(\phi+\theta)}](\Lambda k)^{\mu},
\ee
where $\theta(\Lambda,k)$ and $\beta(\Lambda, k)$ are Little Group rotations and translations, respectively. 
It is then easy to see that each of the soft factor phases satisfy 
\be
e^{-i\rho\frac{\epsilon(k,\phi)\cdot p}{k\cdot p}} = e^{-i\rho\frac{\Lambda \epsilon(k,\phi)\cdot \Lambda p}{\Lambda k \cdot \Lambda p}} = e^{-i\rho Re[\sqrt{2}\beta e^{-i(\phi+\theta)}]} e^{-i\rho \frac{\epsilon(\Lambda k, \phi+\theta) \cdot \Lambda p}{\Lambda k \cdot \Lambda p}}, 
\ee
which is the required covariance condition \eqref{eq:singleCovariance}.
The factor $e^{-i\rho Re[\sqrt{2}\beta e^{-i(\phi+\theta)}]}$ is precisely the appropriate Little Group 
factor derived from $(D_{\phi\phi'}[W(\Lambda, k)])^*$.

In the spin basis, the Lorentz covariance condition \eqref{COVARIANCE} is 
\be
A(\{ k,n \}, \{ p_i \}) = \sum_{n'}(D_{n n'}[W(\Lambda, k)])^* A(\{ \Lambda k,n' \}, \{ \Lambda p_i \}), \label{eq:SingleSpinBasisCovariance}
\ee
which follows from the Bessel function addition theorem upon substitution of \eqref{Soft_spinBasis} and use
of \eqref{eq:Ddefinition}.

What about the intuition that high-spin components of a CSP should lead to trouble?  
Expanding the soft factor \eqref{CSRsoft} as a Taylor series in $\rho$ shows that it contains terms involving $j$ 
powers of $\epsilon_\pm$, for every $j$, mimicking high helicities.  But these high-helicity terms are not arbitrary --- they combine into an overall phase.  
The non-covariant transformation property of $\epsilon$ (the bane of the helicity soft factors) simply re-phases the CSP soft factor 
under Lorentz transformations --- \emph{precisely} the re-phasing needed to account for little-group translations!

Indeed, a striking difference between CSP and helicity amplitudes is this: 
while helicity amplitudes are completely Lorentz invariant (up to a phase), CSP amplitudes in the spin
basis are not --- nor should they be!  
A helicity particle's state is labeled only by its Lorentz-invariant helicity, so the scattering amplitudes must be Lorentz-invariant.  
In contrast, polarization amplitudes for massive spin-1/2 (or higher) particles are not Lorentz-invariant, because the \emph{states} transform non-trivially under boosts.  Likewise, the non-tensor $\epsilon$ appearing in CSP amplitudes 
precisely tracks this transformation of the states (see Figure \ref{fig:spinAngle}), and encodes the non-trivial Lorentz {\it covariance} properties of the amplitude.

What should be true of massive-particle and CSP amplitudes is that the sum over final-state polarizations of $|A|^2$ be 
Lorentz-invariant.  It is instructive to see how this is ensured in an amplitude (motivated by an ``inverse soft factor'' construction) of the form
\be
A(\{k,\phi\},p\dots)= A_0(p\dots) \times \sum_i \frac{1}{\mp 2 k.p_i+ i\epsilon} s_i(\{k,\phi\},p_i)_\pm,
\ee
where each term has a different $\epsilon$-dependent phase, and $A_0$ is a continuation of the $p$-particle amplitude to off-shell momenta.  
There is an overall phase that is not Lorentz invariant\footnote{In fact, the overall phase changes when we 
change the standard boost $B_k$ or our labelling of states. The same is also true of helicity amplitudes.}, 
but the phase \emph{difference} between individual terms $s_i$ and $s_j$ is
\be
\rho \Im\left[e^{i\phi} \left(\frac{{\epsilon^+}.p_i}{k.p_i} - \frac{{\epsilon^+}.p_j}{k.p_j}\right)\right]
= \rho \Im\left[e^{-i\phi} \frac{f^+_{\mu\nu} p_j^\mu p_i^\nu}{k.p_ik.p_j}\right],
\ee
where $f^+_{\mu\nu} \equiv k_\nu \epsilon^+_\mu - k_\mu\epsilon^+_\nu$ is Lorentz-invariant.  
The one overall Lorentz-non-invariant phase drops out of $|A_\phi|^2$ for each $\phi$ state, so that the polarization sum $\int \frac{d\phi}{2\pi}
|A_\phi|^2$ is also Lorentz invariant as required.  The separate Lorentz-invariance of each $|A_\phi|^2$
(before integrating over $\phi$) is a nice feature of the $\phi$ basis, but by no means necessary.
By contrast, individual terms $|A(\{k,n\},p\dots)|^2$ in the spin basis are \emph{not}
Lorentz-invariant, but the invariance of the sum over $n$ follows from the Bessel-function 
completeness identity  $\sum_{n=-\infty}^{\infty} J_n(z)^2 = 1$.

\subsection{Analytic Structure}\label{analytic}
CSP emission and absorption amplitudes constructed from the soft factors \eqref{CSRsoft} are analytic functions of momenta. The non-analyticities are at isolated points, just like ordinary 
tree level scattering amplitudes, but the detailed structure is unfamiliar. In particular, the phase 
of $s_i$ diverges as $k.p_i \rightarrow 0$, which results in an essential singularity at $k.p_i = 0$.  
Similarly, the Bessel-function $n$-basis soft factors \eqref{Soft_spinBasis} vanish as $k.p_i \rightarrow 0$ and are elsewhere analytic in $\epsilon_\pm$, $k$, and $p_i$. Does this new type of singularity introduce any physical pathology?  We should regard it with great skepticism but not contempt --- after all, although multiple poles are almost always problematic, the well-loved $e^{i p.x}$ has an essential singularity at $p.x =\infty$.
Three potential concerns are finiteness of cross-sections in perturbation theory, 
more general problems with analyticity and causality of particle-propagation,
and unitarity of the $S$-matrix. 
It remains an open question whether a full CSP theory can satisfy these conditions, but we can already address 
them in the limited arenas of soft CSP amplitudes and ansatz amplitudes 
built by sewing soft factors together.

A recurring theme will be that, when restricted to physical (real) momenta, the $s_i$ are actually 
bounded functions near their essential singularities.  
As such, there is no need to regulate the $1/k.p_i$ in the phase, and we can consistently keep 
this singular point on the real axis.  It may also be useful to think of these singularities as being regulated 
according to the principal value prescription $1/a\rightarrow \frac{1}{2}(\frac{1}{a+i\epsilon}+\frac{1}{a-i\epsilon})$, 
which is everywhere real on the real line.  

\begin{figure}[!htbp]
\includegraphics[width=0.85\columnwidth]{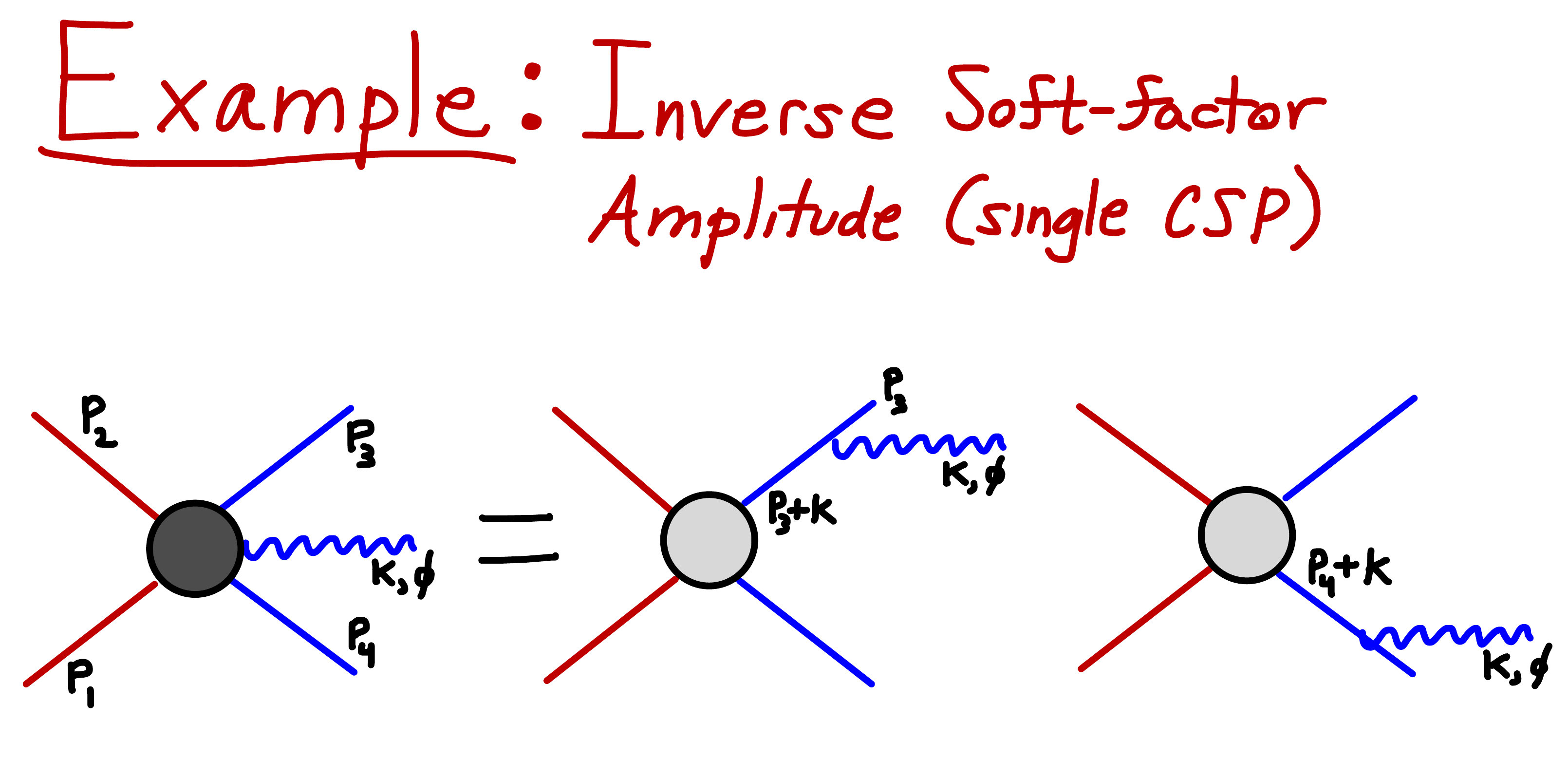}
\caption{The construction of a candidate on-shell CSP amplitude is illustrated above. 
CSPs are attached to a parent amplitude $A_4(p_1,p_2; p_3,p_4)$ using the CSP soft factors, 
with appropriate matter propagators included. This example presumes that only the outgoing 
matter legs couple to the CSP. The final result is the 5-point amplitude $A(p_1,p_2; p_3,p_4,\{k, \phi \})$
used below as an example to investigate certain aspects of CSP interactions. 
\label{fig:single-CSP}}
\end{figure}

As an example, we consider a candidate CSP scattering amplitude built from a soft factor
attached to external scalar legs, as illustrated in Figure \ref{fig:single-CSP}.
We start with a parent amplitude $A_4(p_1,p_2; p_3,p_4)$
of interacting scalar particles (with mass $m$), with the legs $p_3$ and $p_4$ coupled to a CSP. 
We assume for simplicity that the
legs $p_1$ and $p_2$ have no CSP couplings, and that no intermediate-CSP states contribute to the amplitude at leading order in couplings.
To form the resulting 5-point amplitude, we continue $A_4$ to off-shell momenta $p_3+k$ or $p_4+k$, add propagator terms, 
and multiply by the soft factors $s(\{k, \phi \}, p)$. For an external scalar leg with momentum $p$, we evaluate the soft factors
at a momentum $\frac{p+p^*}{2}$, with $p^*=p+k$ in order to symmetrize the soft factor with respect to external matter legs (we note that using $p$, $p^*$, or $p+p^*/2$ yields the same soft factor). 
For the process illustrated in Figure \ref{fig:single-CSP}, this yields
\bea
A(p_1,p_2; p_3,p_4,\{k, \phi \}) &=& A_4(p_1,p_2; p_3+k,p_4)\frac{g_3}{(p_3+k)^2+i\epsilon}s(\{k, \phi \}, p_3)_{-} \nonumber \\
&&+ A_4(p_1,p_2; p_3,p_4+k)\frac{g_4}{(p_4+k)^2+i\epsilon}s(\{k, \phi \}, p_4)_{-}, \label{eq:singleCSP}
\eea
where $g_{3,4}$ are CSP coupling coefficients. 
By construction this amplitude has all the necessary pole contributions required by unitarity (in the matter fields), 
and is Lorentz and Little Group covariant. 
Fourier transforming in $\phi$, we can write down single emission amplitudes in the ``spin'' basis -- this amounts to replacing
$s(\{k, \phi \}, p)_-$ with $s(\{k, n \}, p)_-$.

%\subsubsection{Cross-Sections}
Despite the tower of polarization states available for emission in a scattering process, 
amplitudes of this form give rise to cross-sections that are completely finite, aside from 
familiar infrared singularities. 
In the angle basis, this result is obvious. Because the soft factors are pure phases, the square of the two terms in 
equation \eqref{eq:singleCSP} is unity, so that the only $\phi$-dependent terms in $|A(p_1,p_2; p_3,p_4,\{k, \phi \})|^2$
are the interference terms. 
As promised in the previous subsection, the interference terms contain $\phi$-dependent factors of the form 
\be
e^{i\rho \sqrt{2}\Im\left[e^{-i\phi} \left(\frac{{\epsilon^+}.p_3}{k.p_3} - \frac{{\epsilon^+}.p_4}{k.p_4}\right)\right]}
= e^{i\rho \sqrt{2} \Im\left[e^{-i\phi} \frac{f^+_{\mu\nu} p_3^\mu p_4^\nu}{k.p_3k.p_4}\right]}.
\ee
The cross section for the scattering reaction illustrated in Figure \ref{fig:single-CSP} is proportional 
to 
\\
$\int \frac{d\phi}{2\pi} |A(p_1,p_2; p_3,p_4,\{k, \phi \})|^2$ 
followed by a suitable integration over phase space. 
The $\phi$ integral is unity for the quadratic terms as they are $\phi$-independent, while the interference terms integrate to  
\be
J_0\left( \rho\left|\frac{f^+_{\mu\nu} p_3^\mu p_4^\nu}{k.p_3k.p_4}\right| \right).
\ee
In integrations over the final-state phase space, the non-interference terms produce the usual soft and collinear 
logarithmic divergence arising from $k\cdot p_{3,4}\rightarrow 0$. The analogous divergence in interference terms is regulated 
by the $J_0$ factor.
Thus, aside from IR divergences of the standard type, the cross section is finite. 
This feature persists in other types of CSP amplitudes, including multi-CSP amplitudes,
as discussed in \cite{Schuster:2013vpr}.

In the spin basis, this result appears rather magical. Use of Bessel function 
completeness relations and addition theorems must be used to reduce infinite sums $\sum_{n,m}J_nJ_m$ into
the same result obtained above. That this all works simply in the angle basis,
and is rather awkward in the spin basis, may explain why these results were not obtained in prior literature. 

%\subsubsection{Analyticity, Crossing, and Unitarity}
Ansatz amplitudes like the above (and the soft factor itself) decompose simply as 
a product of a wavefunction that carries Little Group indices and a scalar function of all momenta: 
\be
A(p_1,p_2; p_3,p_4,\{k, \phi \}) = \int d^4\eta \ \psi(\{k,\phi \}, \eta) \ M(\eta,k,p_1,p_2; p_3,p_4).
\ee
Here  $\psi(\{k,\phi \}, \eta)$ is the singular covariant wavefunction \eqref{planewave} with $f(r)=1/r$,
and $M(\eta,k,p_1,p_2; p_3,p_4)$ is a scalar function built out of the $A_4$ sub-amplitude,
\be
M(\eta,k,p_1,p_2; p_3,p_4) = \sum_i e^{i\eta \cdot p_i} A_4(...,p_i+k,...)\frac{g_i}{(p_i+k)^2+i\epsilon}.
\ee
The function $M$ can be analytically continued to complex momentum. This decomposition is similar to the standard 
spinorial amplitude decomposition often used when working with particles of non-trivial spin \cite{Barut,Eden}.

The amplitudes above also satisfy a crossing relation. 
Recall that incoming and outgoing CSP soft factors are related by complex conjugation. 
Thus, an absorption amplitude $A(\{k, \phi \},p_1,p_2; p_3,p_4)$, constructed analogously 
to the amplitude \eqref{eq:singleCSP}, is
\bea
A(\{k, \phi \},p_1,p_2; p_3,p_4) &=& A_4(p_1,p_2; p_3-k,p_4)\frac{g_3}{(p_3-k)^2+i\epsilon}s(\{k, \phi \}, p_3)_{+} \nonumber \\
&&+ A_4(p_1,p_2; p_3,p_4-k)\frac{g_4}{(p_4-k)^2+i\epsilon}s(\{k, \phi \}, p_4)_{+}.
\eea
We define the analytic continuation of $\epsilon_{\pm}(k)$ from $k^0>0$ to $k^0<0$ in the usual way,
so that $\epsilon_{\pm}(-k)=-\epsilon_{\mp}(k)$. 
Having done that, we obtain the crossing relation 
\be
A(\{-k, \phi \},p_1,p_2; p_3,p_4) = A(p_1,p_2; p_3,p_4,\{k, \phi+\pi \})
\ee
which is indeed satisfied given our soft factors \eqref{CSRsoft} so long as $f_i^+(p_i.k) = f_i^-(-p_i.k) = f_i^+(-p_i.k)^*$. The left-hand side
is understood as the analytic continuation from $k$ to $-k$, while the right-hand side
has been PT conjugated, sending $\phi\rightarrow \phi+\pi$ (see \eqref{eq:PT}). 
An overall phase associated with PT, that we can set to unity by convention, has been dropped. 
In theories where particles are there own anti-particles, as we assume here, 
crossing relates the analytic continuation of the absorption amplitude from $k$ to $-k$ 
to emission amplitudes for the PT-conjugate of the crossed state (see \cite{Barut,Eden}),
which is precisely what we have. 

Satisfied that our emission and absorption amplitudes are related by a familiar crossing relation,
we next look at the optical theorem constraints. With $X$ and $Y$ denoting the incoming and 
outgoing matter particles, respectively, the optical theorem applied to the soft emission amplitudes is
\be
T_{X \rightarrow Y+{k_\phi}} - (T_{Y+{k_\phi} \rightarrow X})^*  = -i \sum_Z T_{X\rightarrow Z} (T_{Y+{k_\phi}\rightarrow Z})^*.
\ee
The concern about unitarity from a multiple-pole or essential singularity would be that the left-hand-side 
would receive a contribution that does not arise from a physical intermediate state $Z$ on the 
right-hand side.  But in fact, the CSP phase factors in the two terms on the left hand side, as well as the 
right hand side, are identical -- the only ``discontinuity'' comes from the propagators' explicit $i\epsilon$ factors 
that \emph{do} correspond to on-shell intermediate states that undergo soft or collinear splitting.  
The absence of new discontinuities relies crucially on the boundedness of the $p.k\rightarrow 0$ limit, so that there is no need to deform the singularity from the real line.  In loop calculations for example, 
we can integrate straight through the singularity in any contour prescription and not pick-up 
any spurious imaginary parts of amplitudes.  Had we replaced the CSP phase with a higher-order pole, it would necessarily have diverged as $p.k\rightarrow 0$ (even at real momenta).  Regulating such a divergence by deforming into the complex plane would have introduced a new unphysical discontinuity. 

While simple tree-level on-shell CSP amplitudes present no obvious conflict with unitarity and 
analyticity, simple scattering amplitudes with ``off-shell'' CSPs could present difficulties --- this will be studied in the simplest type of ``off-shell'' CSP-exchange 
process in \cite{Schuster:2013vpr}. These amplitudes will continue to be consistent with analyticity, 
though again possessing structure that will make it difficult to assess whether new problems arise. 
While we have investigated numerous examples, we have been unable to successfully ``bootstrap'' our 
way to a complete CSP theory. Instead, we take the soft factors and their helicity correspondence as a guide in searching for a field theory 
(or other type of) description of CSP dynamics.

%%%%%%%%%%%%%%%%%%%%%%%%%%%%%%%
\section{Conclusions and Future Directions}\label{sec:conclusion}
In this paper, we have reported on several encouraging results that we believe open the way to 
new and potentially productive lines of investigation. 
We have derived the most general wavefunctions for (single-valued) CSPs. 
We have discovered a family of new wave equations, the solutions of which play an important role in describing CSP interactions. 
Our new wave equations make contact at $\rho=0$ with standard descriptions of massless particles of arbitrary helicity. Finally, we have discovered a class of CSP soft factors that can readily be used to construct candidate CSP scattering amplitudes
consistent with unitarity that yield finite cross-sections.  

One unfamiliar feature of the soft factors is an essential singularity at soft and/or collinear configurations.  
This singularity is bounded for all real momenta (taking the form of a rapidly oscillating phase or a damped Bessel function), 
 and does not pose any obvious physical problem, nor any sharp obstruction to analytic continuation for the simple
 examples we have studied. However, it is not clear if this new structure can arise in a more complete theory with manifestly local interactions.  
 A small but perhaps significant difference between our treatment and most of the continuous-spin particle literature 
is our use of the ``angle basis'' (diagonalizing Little Group translations) rather than the ``spin basis'' (diagonalizing Little Group rotations).  
Many results that rely on Bessel Function identities in the spin basis are trivially obtained in the angle basis.  

In a companion paper \cite{Schuster:2013vpr}, we exhibit a remarkable property of the soft factors presented here: 
at energies higher than $\rho$, the soft-emission amplitudes in the ``spin'' basis become hierarchical, with spin-0, $\pm 1$, or $\pm 2$ 
amplitudes well approximated by the corresponding helicity amplitudes, and emission of all other spin states suppressed by $\rho v/E$ for 
characteristic velocities $v$ and CSP energy $E$.  We call this behavior \emph{helicity correspondence} and conjecture 
that there exist CSP theories where this correspondence extends to all scattering amplitudes.  
An important consequence of the helicity correspondence is that it naturally resolves the oldest objection to physical CSP theories, 
raised by Wigner --- that, because they contain infinitely many degrees of freedom, the vacuum has infinite heat capacity density. 
Though this is formally true, it may not be problematic for any realistic physical system.  After all, real systems are only \emph{approximately} 
in thermodynamic equilibrium, and only for a finite time.  If additional CSP states are sufficiently weakly coupled, then the timescale for 
quasi-thermal systems to dissipate energy into those states is long enough to be physically irrelevant.  
We expand on this argument quantitatively in \cite{SchusterToro:ph}.

The largest question raised by the existence of CSP soft factors is what theory (if any) gives rise to them.  It may be, of course, that like the 
Veneziano amplitude, interacting CSPs arise only in theories with additional structure \emph{beyond} local fields.  If so, it would be very 
interesting to find this structure!  We have, however, found indications that a \emph{gauge theory} of CSPs coupled to a background current 
does exist -- this is the topic of our upcoming paper \cite{Schuster:2013pta} -- though a definitive understanding of interactions and their locality
properties is still lacking. 
This underscores the value of continuing to apply on-shell methods to resolve several important open problems:  

\textbf{CSP Self-Interactions:}
The soft factors proposed in this paper are exactly covariant (even away from the soft limit) for scalar matter and can be generalized to obtain covariant interactions at finite CSP momentum with matter of non-trivial spin.
But they do not readily generalize to multi-CSP interactions, which involve off-shell CSP propagators and could have a more intricate phase structure.  Probing the consistency and structure of CSP self-interactions is especially interesting in light of the helicity correspondence, which suggests that CSPs could have gravity-like interactions with matter \cite{Schuster:2013vpr}.  The consistency of such a theory surely requires CSP self-interactions as well.  One approach to studying these interactions, familiar from modern unitarity-based methods in gauge theories and gravity, is to define on-shell, momentum-conserving three-particle ``amplitudes'' at complex momentum.  
These ``amplitudes'' are, however, supported on very 
degenerate collinear momentum configurations that push the naive CSP soft factors to their ``singular'' limit. 
CSP self-interactions may instead be easiest to investigate by looking for true real-momentum amplitudes, 
starting at four-particle level -- these are the simplest amplitudes with non-degenerate momentum-space support.  

\textbf{Multi-CSP Couplings to Matter and Fermionic CSPs:}
We have argued that the soft factors \eqref{CSRsoft} and \eqref{Soft_spinBasis} are the only ones consistent with both Lorentz-invariance (which implies a simple relationship with wavefunctions) and cluster decomposition (which requires momentum-space smoothness of amplitudes).  Amplitudes involving two CSPs must transform like a two-particle state, and could be of a form more general than \eqref{integralContract}.  For example, instead of simply evaluating each $\eta$-space wavefunction at $\eta=p$ for some particle momentum $p$, one can also integrate  two such functions against each other.  A classification of such invariants would be very useful in studying the physics of CSP interactions with matter, and may also make contact with the case of double-valued CSPs.

\textbf{Continuation to Complex Momenta, Factorization, and Unitarity:}
It would be interesting to investigate the analytic continuation of tree-level CSP soft factors and associated 
amplitudes to complex momentum. Likewise, factorization and generalized unitarity would be interesting 
to understand in this setting. 
One of our early goals was to define a tree-level interacting CSP theory by recursive techniques (e.g. BCFW-style recursion relations \cite{Britto:2004ap,Britto:2005fq}). Though this is not out of the question, some difficulties must be addressed. 
On the one hand, the structure of the CSP soft factors associated with $\rho\neq 0$ posses no problem 
with BCFW shifts, and is in fact well-behaved in the limit of large shifts. 
However, as CSP amplitudes are not rational functions of momentum anymore, contour integrals of 
BCFW-deformed amplitudes are less directly related to a sum of residues that can be computed in a straightforward fashion.  
That said, the functions appearing in the CSP soft factors (Bessel functions in the spin basis or phases in the angle basis) have a very restricted structure, so it seems likely that recursion relations exist -- these should be found.  

%%%%%%%%%%%%%%%%%%%%%%%%
\section*{Acknowledgments}

We thank 
Haipeng An,
Nima Arkani-Hamed,
Cliff Burgess, 
Freddy Cachazo,
Yasunori Nomura,
Maxim Pospelov, 
Yanwen Shang,
Carlos Tamarit,
Mark Wise,
Edward Witten, 
and Itay Yavin
for helpful discussions on various aspects of this physics.
We additionally thank Freddy Cachazo and Jared Kaplan for useful feedback on the manuscript. 
Research at the Perimeter Institute is supported in part by the Government of Canada through NSERC and by the Province of Ontario through MEDT.  
	
%%%%%%%%%%%%%%%%%%%%%%%%%%%%%%%%%%%%%%%%%%
\appendix
\section{Conventions and Identities}\label{App:frame}
This appendix provides a self-contained introduction to the Little Group of the Poincare group in 3+1 dimensions, with a focus on interpolating between the massive and massless Little Groups.  

We work in the mostly-negative metric, with $\epsilon^{0123} = -\epsilon_{0123} = +1$.
Consider a null or timelike vector $k$ with $k^2=M^2$.  For any null $q$ such that $q.k\neq 0$, we can define a null vector
$k_\flat = k - \f{M^2}{2 k.q} q$ and polarizations $\epsilon_{\pm}$ and $\epsilon_0$ satisfying 
\bea
\epsilon_\pm^2 = 0;\qquad \epsilon_\pm.(k,k_\flat,q) = 0; \qquad \epsilon_+.\epsilon_- = -1; \qquad \epsilon_- = \epsilon_+^* \\
k_\flat.q = k.q \\
\epsilon_0.(k,\epsilon_\pm) = 0 ; \qquad \epsilon_0^2 = -1.
\eea
The unique vector $\epsilon_0$ satisfying these conditions is given by
\be
\epsilon_0 = \f{1}{M}( k - \f{M^2}{k.q} q )  = \f{1}{M}( k_\flat - \f{M^2}{2 k.q} q ).
\ee
As $M^2\rightarrow 0$, $\epsilon_0$ becomes ill-defined (the normalization tends to infinity) but $M\epsilon_0$ approaches $p$.  

By convention we will always take $k$ and $q$ to have \emph{positive} time components, and  $\epsilon^+$ to be such that a receiver looking back toward the origin from the $+{\mathbf k}$ direction sees $\epsilon^+ e^{-i\omega t}$ rotating counter-clockwise.  For example, one such standard basis is
\be
k_\flat^\mu = (k, 0, 0, k)\quad
q^\mu = (\varq,0,0,-\varq) \quad
\epsilon_{\pm}^{\mu} = (0, 1, \pm i, 0)/\sqrt{2}.
\ee

The conditions above, together with the choice of $k$ and $\epsilon_+$, fully specify the frame (a choice of $k$ and $q$ fixes the frame up to the phase of $\epsilon_+$).  

Since the frame forms a basis for all Lorentz vectors, we can decompose the invariants $\epsilon^{\mu\nu\rho\sigma}$ and $g^{\mu\nu}$ as 
\be
\epsilon^{\mu\nu\rho\sigma} = \f{-i}{k.q} \left[ \epsilon_+^\mu \epsilon_-^\nu k_\flat^\rho q^\sigma \pm perms \right],
\ee
where the sum is over all permutations of indices, with alternating signs, and 
\be
g^{\mu\nu} = -(\ep_+^\mu\ep_-^\nu + \ep_-^\mu\ep_+^\nu) + \f{1}{k.q} (k_\flat^\mu q^\nu + q^\mu k_\flat^\nu).
\ee

\section{The Little Group Generators and Their Actions}\label{App:LG}
\subsection{Massive Case}
Using the results of \ref{App:frame} we can simplify the results of contracting $\epsilon_\pm$ and $\epsilon_0$ into $w^\mu$.  For massive $k^2 = M^2$ we find  
\bea
W_0 \equiv \epsilon_0.W & = & \frac{1}{M}(k - \f{M^2}{k.q} q).W = -\f{M}{k.q} q.W \\ 
 & = & -\f{M}{k.q} \f{1}{2} q_\mu k_\nu \epsilon^{\mu\nu\rho\sigma} J_{\rho\sigma} \\
& = & -\f{M}{k.q}\f{1}{2} q_\mu k_\nu \pfrac{-i}{k.q} \left(k_\flat^\mu q^\nu \epsilon_+^\rho \epsilon_-^\sigma - (\rho\leftrightarrow\sigma)+ \dots\right) \\
& = & i M \epsilon_+^\rho \epsilon_-^\sigma J_{\rho\sigma}.\label{eq:W0massive}
\eea
In the third line, $\dots$ denotes terms in the expansion of $\epsilon^{\mu\nu\rho\sigma}$ which vanish when contracted with $q_\mu k_\nu$.
Moreover, since 
\be
{(J_{\rho\sigma})^\alpha}_\beta = i (\delta^{\alpha}_\rho g_{\sigma\beta} - \delta^{\alpha}_\sigma g_{\rho\beta}),
\ee
we find 
\be
{(W_0)^\alpha}_{\beta} = M (\epsilon_{-}^\alpha\epsilon_{+\beta} - \epsilon_+^{\alpha}\epsilon_{-\beta}).
\ee
Similar computations for the $\epsilon_{\pm}$ yield
\bea
%W_0 &\equiv& \epsilon_0.W = -\f{M}{k.q} q.W = i M \epsilon_+^\rho \epsilon_-^\sigma J_{\rho\sigma} \qquad \quad
%{(W_0)^\alpha}_{\beta} = M (\epsilon^{-\alpha}\epsilon^+_\beta - \epsilon^{+\alpha}\epsilon^-_\beta) \\
W_+ & \equiv & \sqrt{2} \epsilon_+.W =   i\sqrt{2} M \epsilon_{0}^{\rho}\epsilon_+^\sigma J_{\rho\sigma} \qquad \qquad \quad
{(W_+)^\alpha}_{\beta} =\sqrt{2} M (\epsilon_{+}^{\alpha}\epsilon_{0\beta} - \epsilon_0^{\alpha}\epsilon_{+\beta}) \\
W_- & \equiv & \sqrt{2} \epsilon_-.W =  - i\sqrt{2} M \epsilon_{0}^{\rho}\epsilon_-^\sigma J_{\rho\sigma} \qquad \qquad
{(W_-)^\alpha}_{\beta} = -\sqrt{2} M (\epsilon_-^{\alpha}\epsilon_{0\beta} - \epsilon_0^{\alpha}\epsilon_{-\beta}.)
\eea
These satisfy the commutation relations
\be
[W_0,W_\pm] = \pm M W_\pm\qquad [W_+,W_-] = 2 M W^0 
\ee
so that $J_{0,\pm} \equiv W_{0,\pm}/M$ satisfy the usual $SO(3)$ commutation relations for massive particles.  
We can decompose $W^\mu$ in terms of these generators as
\be
W^\mu = -\f{1}{\sqrt{2}}\left(\epsilon_-^\mu W^+ + \epsilon_+^\mu W^- \right) -\epsilon_0^\mu W^0
\ee
so that
\be
W^2 = -\f{1}{2} (W^+W^- + W^-W^+) - (W^0)^2 = -M^2 J^2.
\ee

\subsection{Massless Limit}
As $M\rightarrow 0$, $M\epsilon^0$ approaches $k$ giving $W_\pm$ well-defined limits, which we call $T_\pm$.  $W_0$ approaches zero, but this is an artifact of the diverging normalization of $\epsilon_0$ -- indeed, the term $\epsilon_0^\mu W^0$ has a finite limit.  This motivates defining a finite, rescaled $\tilde W_0 \equiv -\f{1}{k.q} q.W$, which for any finite $M$ is nothing but $W_0/M$.  We denote the massless limit of $\tilde W_0$ as $R$.  For the massless case, we then have
\bea
T_\pm &=& \pm i \sqrt{2} k^\rho \epsilon_\pm^{\sigma} J_{\rho\sigma} \\
R & =& i \epsilon_+^{\rho} \epsilon_-^{\sigma} J_{\rho\sigma}.
\eea
in terms of which 
\be
W^\mu = -\f{1}{\sqrt{2}}\left(\epsilon_-^\mu T_+ + \epsilon_+^\mu T_- \right) -k^\mu \tilde R, \qquad W^2 = - T_+T_-.
\ee
These satisfy the commutation relations
\be
[T_+,T_-] = 0 \qquad [R, T_\pm ] = \pm T_\pm
\ee
of the two-dimensional Euclidean group $ISO(2)$.

%\end{document}

%%%%%%%%%%%%%%%%%%%%
\section{Continuous-Spin Wave Equations}
Our goal in this appendix is to derive the most general solution to \eqref{Reom}-\eqref{T+eom}. 
Solutions to this equation can be used to derive ansatz soft factors for particle emission, and so their
role is central. 
As described in Appendix \ref{App:LG}, the infinitesimal form of a Little Group transformation
acting on a four-vector is $\Lambda = 1+ \frac{i}{\sqrt{2}}\beta w_{-} + \frac{i}{\sqrt{2}}\beta^*w_{+}$ for LG translations
and  $\Lambda = 1-i \theta w_{r}$ for rotations.  
Using \eqref{LGgenAction}, the rotation and translation equations for $\psi(\{p,\phi \},\eta)$ are 
\bea
-i\left( \eta.\epsilon_{-} \epsilon_{+}.\partial_{\eta}-\eta.\epsilon_{+} \epsilon_{-}.\partial_{\eta} \right) \psi &=& \partial_{\phi} \psi \label{eq:appendixLGa} \\
-\left( \eta.\epsilon_{-}p.\partial_{\eta} - \eta.p\epsilon_{-}.\partial_{\eta} \right) \psi &=& \frac{\rho}{\sqrt{2}} e^{-i\phi} \psi \label{eq:appendixLGb} \\
\left( \eta.\epsilon_{+}p.\partial_{\eta} - \eta.p\epsilon_{+}.\partial_{\eta} \right) \psi  &=& \frac{\rho}{\sqrt{2}} e^{i\phi} \psi \label{eq:appendixLGc}
\eea
These equations are homogeneous in $\eta$, as we'd expect 
for an equation picking out Lorentz covariant wavefunctions for representations of the Poincare group. 

To find the most general solutions to the Little Group covariance equations, we start 
by noting that the Little Group operators annihilate $\eta.p$ and $\eta^2$,
while the rotation operator also annihilates $\eta.q$. 
Note that the rotation operator is diagonal in terms of $\eta.\epsilon_{\pm}$,
\be
i\eta.(w_r).\partial_{\eta} (\eta.\epsilon_{\pm}) = \pm i (\eta.\epsilon_{\pm}).
\ee
Thus, the most general solution to the rotation equation can be written as 
\be
\psi(\{p,\phi \},\eta) = g(z,\bar z, a, b),
\ee
where $z$ is the complex variable $z\equiv \eta.\epsilon_{+}e^{-i\phi}$, $a\equiv \eta.q / p.q$, and $b\equiv \eta\cdot p$.
The translation equations now reduce to 
\bea
-\left( \bar{z}\partial_a + b \partial_z \right) g(z,\bar z, a, b) &=& \frac{\rho}{\sqrt{2}}g(z,\bar z, a, b) \label{eq:transA}\\
\left( z\partial_a + b \partial_{\bar{z}} \right) g(z,\bar z, a, b) &=& \frac{\rho}{\sqrt{2}}g(z, \bar z, a, b). \label{eq:transB}
\eea
We must consider separately two families of solutions: those with 
$f\sim \delta(\eta.p)$ (singular) and those with smooth dependence on $\eta.p$.

\subsection{Singular Solutions and Wave Equations}
In the singular (in $p\cdot \eta$) case, the translation equations simplify to 
$z\partial_a g = - \bar z \partial_a g = \frac{\rho}{\sqrt 2} g$.  For non-zero $\rho$, this can only be satisfied by $g$ localized on $\bar{z}=-z$,
so that the most general solution is 
\bea
\psi(\{p,\phi \},\eta) &=& \delta(\eta.p)f(\eta^2)\delta(z+\bar{z})e^{\frac{\rho}{\sqrt{2}}\frac{a}{z}} h(z) \nonumber \\
&=& \delta(\eta.p)\hat f(\eta.\epsilon_+ e^{-i\phi})\delta(\eta.\epsilon_{+}e^{-i\phi}+\eta.\epsilon_{-}e^{i\phi})e^{\frac{\rho}{\sqrt{2}}\frac{\eta.q e^{i\phi}}{\eta.\epsilon_{+}p.q}}, 
\eea
where the functions $f$, $h$, and $\hat f$ are arbitrary. In obtaining the second line, we have used the fact that, on the support of 
$\eta.p=0$, $\eta^2=-2|z|^2=2z^2$ to combine $f(\eta^2)$ and $h(z)$ into a single function $\hat f(z)$.  
This may be written more nicely as
\be
\psi(\{p,\phi,f\},\eta)=\int dr f(r) \int d\tau \delta^{4}(\eta-r\epsilon(p\phi)-r \tau p)e^{-i\tau\rho}, \label{planewaveAPP}
\ee
where $f(r)$ is arbitrary (and distinct from our previous $f(\eta^2)$) and 
\be
\epsilon(p\phi) \equiv \frac{i}{\sqrt{2}}( \epsilon_{+}e^{-i\phi}-\epsilon_{-}e^{i\phi}).\label{epphi}
\ee
We note that this definition of $\epsilon(p\phi)$ is rephased by $\pi/2$ relative to the usual convention: 
here $\epsilon(\phi=0)  =  (i \epsilon_+ - i \epsilon_-)/\sqrt{2}$, rather than $(\epsilon_++\epsilon_-)/\sqrt{2}$. 
However the combination $\epsilon(p,\phi)$ as defined in \eqref{epphi} will be ubiquitous in our calculations.  

On the support of the $\delta$-function in \eqref{planewaveAPP}, $\eta^2+r^2=0$, but the wavefunction's dependence on $r$ is dictated by the unconstrained function $f$. Since $\epsilon(p,\phi+\pi)=-\epsilon(p,\phi)$, 
the wavefunctions labelled by $\{\phi,f(r)\}$ and $\{\phi+\pi,f(-r)\}$ are equal for any $f$.  It is therefore natural to restrict our attention to $f(r)$ supported on $r>0$, for which 
the wavefunctions \eqref{planewaveAPP} satisfy
\bea
\eta.p \psi &=& 0 \label{wigner1}\\ 
\left( - i \sqrt{-\eta^2} p.\partial_{\eta} + \rho \right) \psi &=& 0. \label{wigner2}
\eea
The second equation can be interpreted as a simplification (and linearization) of 
\be
\left( W^2 + \rho^2 \right) \psi = 0,
\ee
on the support of $\eta.p \psi =0$. These two equations together imply $p^2\psi=0$.
If we consider $f(r)=\delta(r-1)$, the resulting wavefunctions 
also satisfy $(\eta^2+1)\psi=0$.  This restricted set of wavefunctions satisfy the Wigner equations 
for continuous spin representations \cite{Wigner:1939cj,Wigner:1947,Bargmann:1946me}.

We stress that different choices of $f(r)$ correspond to different conditions on the $\eta^2$-dependence, which 
can for example be specified by an $\eta$-dependent Lorentz covariant equation. 
For example, solutions to the fully homogenous Lorentz covariant equations,
\bea
p^2 \psi &=& 0 \\
\eta.p \psi &=& 0 \\
 \eta \cdot \partial_{\eta} \psi &=& 0 \\
\left( - i \sqrt{-\eta^2} p.\partial_{\eta} + \rho \right) \psi &=& 0, 
\eea
are Little Group covariant wavefunctions. These rather simple homogenous equations are new. 

\subsection{Smooth in $p \cdot \eta$ Solutions and Wave Equations}
Taking $z$ times \eqref{eq:transA}  and $\bar{z}$ times \eqref{eq:transB}, we can remove the $a$-derivative dependence from 
the sum of the two translation equations, giving 
\be
\left( -z\partial_z + \bar{z}\partial_{\bar z} \right) g(z,\bar{z},a,b) = \frac{\rho(z+\bar{z})}{\sqrt{2}b} g(z,\bar{z},a,b).
\ee 
This is solved by 
\be
h(b,|z|^2,a)e^{\frac{\rho}{\sqrt{2}b}(\bar{z}-z)}
\ee
with $h$ arbitrary. The second translation equation now becomes 
\be
z\partial_a h(b,|z|^2,a) = 0,
\ee
so that solutions with support at non-zero $z$ have no $a$-dependence. 
We can trade $b$ and $|z|^2$ for $\eta\cdot p$ and $\eta^2$, which allows 
us to write the general solution of \eqref{eq:appendixLGa}-\eqref{eq:appendixLGc} as 
\bea
\psi(\{p,\phi\},\eta) &=& f(\eta.p,\eta^2) e^{i\rho \frac{\eta.\epsilon(p\phi)}{\eta.p}} \nonumber \\
&=&   f(\eta.p,\eta^2) e^{-i\rho\sqrt{2} \Im \left( \frac{\eta.\epsilon_{+}e^{-i\phi}}{\eta.p} \right) },
\eea
with $f(\eta.p,\eta^2)$ an arbitrary function. 

Compared to the singular case, wavefunctions of this form can arise as solutions to a broader class of wave equations,
essentially because two conditions can be imposed to pick out the $\eta\cdot p$ and $\eta^2$ dependence in $f$. 
One particulalrly useful class of wave equations for CSPs that we use is 
\bea
p^2 \psi &=& 0 \\
p\cdot \partial_{\eta} \psi &=& 0 \\
 \eta \cdot \partial_{\eta} \psi &=& n \psi \\
\left( - (p\cdot \eta)^2\partial_{\eta}^2 + \rho^2 \right) \psi &=& 0.
\eea
When $\rho=0$, this coincides with the transverse and traceless gauge-fixed form of the Fronsdal equation for massless particles 
of helicity $n$. Other interesting wave equations can be used, as will be discussed in \cite{Schuster:2013pta}. 

%%%%%%%%%%%%%%%%%%%%%%%
\bibliography{CSP_kinematics}
%%%%%%%%%%%f%%%%%%%%%%%%
\end{document}